\documentclass[twocolumn]{aastex62}
\usepackage{amsmath,amstext}
\usepackage[figure,figure*]{hypcap}
\usepackage{newtxmath} 

\usepackage{mymacros}
\usepackage{aas_macros}
\usepackage{xspace}

\graphicspath{{./}{figures/}}

\shorttitle{M82 X-2: Transient pulsations, spin-down and a glitch}
\shortauthors{Bachetti et al.}

\begin{document}

\title{All at once: transient pulsations, spin-down and a glitch from the Pulsating Ultraluminous X-ray Source M82~X-2}

\author[0000-0002-4576-9337]{Matteo Bachetti}®
\altaffiliation{Fulbright Visiting Scholar}
\affiliation{INAF-Osservatorio Astronomico di Cagliari, via della Scienza 5, I-09047 Selargius, Italy}
\affiliation{Space Radiation Laboratory, Caltech, 1200 E California Blvd, Pasadena, CA 91125}
\email{matteo.bachetti@inaf.it}

\author{Thomas J. Maccarone}
\affiliation{Department of Physics and Astronomy, Texas Tech University, Lubbock, TX, USA}

\author{Murray Brightman}
\affiliation{Space Radiation Laboratory, Caltech, 1200 E California Blvd, Pasadena, CA 91125}

\author{McKinley C. Brumback}
\affiliation{Department of Physics \& Astronomy, Dartmouth College, 6127 Wilder Laboratory, Hanover, NH 03755, USA}

\author{Felix F\"urst}
\affiliation{European Space Astronomy Centre (ESA/ESAC), Operations Department, Villanueva de la Canada Madrid, Spain}

\author{Fiona A. Harrison}
\affiliation{Space Radiation Laboratory, Caltech, 1200 E California Blvd, Pasadena, CA 91125}

\author{Marianne Heida}
\affiliation{Space Radiation Laboratory, Caltech, 1200 E California Blvd, Pasadena, CA 91125}

\author{Gian Luca Israel}
\affiliation{Osservatorio Astronomico di Roma, INAF, via Frascati 33, I-00078 Monte Porzio Catone, Italy}

\author{Matthew J. Middleton}
\affiliation{Department of Physics and Astronomy, University of Southampton, Highfield, Southampton SO17 1BJ, UK}

\author{John A. Tomsick}
\affiliation{Space Sciences Laboratory, University of California, Berkeley, 7 Gauss Way, Berkeley, CA 94720, USA}

\author{Natalie A. Webb}
\affiliation{IRAP, Universit\'é de Toulouse, CNRS, UPS, CNES, 9 Avenue du Colonel Roche, BP 44346, 31028 Toulouse Cedex 4, France}

\author{Dominic J. Walton}
\affiliation{Institute of Astronomy, University of Cambridge, Madingley Road, Cambridge CB3 0HA, UK}

\begin{abstract}
\Mtwo is the first pulsating ultraluminous X-ray source (PULX) to be identified. 
Since the discovery in 2014, \nustar has observed the M82 field 15 times throughout 2015 and 2016.
In this paper, we report the results of pulsation searches in all these datasets, and find only one new detection.
This new detection allows us to refine the orbital period of the source and measure an average spin-down rate between 2014 and 2016 of $\sim -6\times 10^{-11}$ Hz/s, which is in contrast to the strong spin-up seen during the 2014 observations, representing the first detection of spin-down in a PULX system.
Thanks to the improved orbital solution allowed by this new detection, we are also able to detect pulsations in additional segments of the original 2014 dataset. 
We find a glitch superimposed on the very strong and variable spin-up already reported -- the first positive glitch identified in a PULX system.
We discuss the new findings in the context of current leading models for PULXs.
\end{abstract}


\section{Introduction}

Ultraluminous X-ray sources (ULX) are off-nuclear point sources with X-ray luminosities exceeding the Eddington limit for a stellar-remnant black hole \citep[see][for a review]{kaaretUltraluminousXRaySources2017}.
In 2014, \citet{Bachetti+14} (hereafter B14) reported pulsations from the known ULX \Mtwo, showing that at least some ULXs are neutron stars. 
With additional observations and timing analysis, more of these objects are being found to pulsate (e.g. 
NGC 5907 X-1: \citealt{israel_accreting_2017}; 
NGC 7793 P13: \citealt{israel_discovery_2017,furst_discovery_2016}; 
NGC 300 X-1: \citealt{carpano_discovery_2018}; 
NGC 1313 X-2: \citealt{sathyaprakashDiscoveryWeakCoherent2019}).
The spectral and variability properties of these pulsating ultraluminous X-ray sources (PULX\footnote{also referred to as ultraluminous x-ray pulsars, ULXP}) are similar to the bulk of the ULX population.
Looking in detail, they tend to have slightly harder spectra and a higher level of long term-variability than average%
\footnote{For a detailed view see \citealt{pintore_pulsator-like_2017,walton_evidence_2018,koliopanos_ulx_2017}}.
This fact, together with the intrinsic difficulties in finding pulsations in these distant sources, means that it is plausible that most of the ULX population is powered by neutron stars\footnote{For deeper investigations of ULX populations, see \citet[][]{middleton_predicting_2017,wiktorowicz_observed_2018}}.

The PULXs found so far have in common periods around 1\,s (with the notable exception of NGC 300 X-1, which is much slower), and strong spin-up during observations and between observations spaced by months to years. 
The spin-up is likely due to the torque of an accretion disk onto the star as observed in many accreting neutron stars in X-ray binary systems \citep[e.g.][]{Bildsten:1997tp,rappaportAccretionTorquesXray1977,pernaSpinupSpindownTransitions2006,burderiOrderChaosSpinup2006}.

Accretion onto magnetized neutron stars happens through the interaction of the accretion disk with their strong magnetic field.
The general framework of accretion onto magnetized objects is described in a number of papers \citep[e.g.][]{GL79a,GL79b,wang_torque_1995} and reviewed in \citet{FrankKingRaine}.
The disk is truncated at a distance \rin (often called the Alfv\'en radius) from the NS, where magnetic stresses equal the ``ram pressure'' from the matter in the disk. 
Therefore, the same \rin can be determined by hugely different magnetic field strengths, provided that they are balanced by different mass accretion rates.
The magnetic field, besides interrupting the disk at \rin, also penetrates the disk beyond the inner radius. 
The interaction between the matter and the magnetic field at a given distance from the NS produces a torque on the star that depends on the relative angular velocity of the matter with respect to the star.
Calling \rco the \textit{corotation} radius, the radius at which the Keplerian angular velocity of the accreting matter equals the angular velocity of the star, the total torque depends on how many field lines thread the disk inside and beyond \rco \citep{wang_torque_1995}.
The observation of spin-up and spin-down in accreting pulsars with plausibly very different magnetic fields and companion stars is usually explained in this theoretical framework with good success.
As a final ingredient in standard accreting pulsars, the X-ray luminosity is considered a proxy of mass accretion rate \citep[following][]{SS73}, as for thin disks we expect a relation $L\propto\Mdot$.

However, the luminosity of PULXs is far higher than the Eddington luminosity for a neutron star, even considering the modifications to the Eddington limit that arise in pulsar accretion columns \citep[e.g.][]{BaskoSunyaev76}.
This has led to a lively theoretical debate, which is still ongoing. 

Because of this super-Eddington accretion, any treatment of PULXs should include a third special radius, called \textit{spherization} radius \rsph. 
In super-Eddington disks, inside \rsph, we cannot assume a standard thin disk configuration. Inside this radius, the disk is swollen and strong winds are ejected due to radiation pressure, limiting the total amount of matter that actually approaches the inner radius and participates in the acceleration of the star \citep{SS73,Poutanen+07}.
Moreover, the observed luminosity itself can be misleading. 
The bolometric luminosity of a super-Eddington disk increases according to 
\begin{equation}
L_{\rm acc} \approx L_{\rm Edd} \left(1 + \ln \mdot\right),
\end{equation}
where $\mdot=\Mdot/\Mdot_{\rm Edd}$.
However, this is probably not the bolometric luminosity measured by the observer, because radiation might be beamed.
\citet{kingMassesBeamingEddington2009} describe a geometry where the origin of the beaming is the wind. 
In their model, the wind angle is proportional to the mass accretion rate, and
this leads to a beaming factor $b\propto\Mdot^{-2}$.
In the end one obtains
\begin{equation}
L_{\rm obs} =L_{acc} / b \propto L_{\rm Edd} \mdot^2 \left(1 + \ln \mdot\right),
\end{equation}

With this in mind, the literature about PULXs has explored two main cases: \rsph$\gtrsim$\rin, and \rsph$<$\rin \citep[for recent examples, see][]{walton_evidence_2018,middletonMagneticFieldM512019,mushtukovTimingPropertiesULX2019}.

If \rsph$>$\rin, the disk becomes locally super-Eddington before reaching the inner radius. 
So, the actual mass accretion rate (the mass that reaches \rin and accretes on the NS) is lower than inferred from the luminosity, because winds eject a large fraction of the matter, and, additionally, the luminosity itself is beamed and overestimated. 
The observation of blueshifted lines reminiscent of strong winds, and various kinds of nebulae around ULXs, supports this hypothesis \citep[e.g.][]{griseOpticalPropertiesUltraluminous2011,pinto_ultrafast_2016}.
However, the bulk of observations of nebulae, wind signatures and broad pulse profiles exclude extreme beaming factors \citep[see review by][]{kaaretUltraluminousXRaySources2017}.

In the opposite case, the disk remains thin until it reaches \rin, and no strong winds are launched.
The effects of the huge accretion rate manifest themselves in the accretion column, giving rise to an optically thick envelope around the star that can in principle suppress high-frequency variability \citep{mushtukovOpticallyThickEnvelopes2017}.
Since the inferred \rin is quite large and the inferred magnetic fields are magnetar-like, this model justifies the high luminosity because, in very high magnetic fields, the Thompson scattering cross section -- one of the key ingredients of the Eddington limit -- is modified, and the Eddington limit itself is much higher \citep{BaskoSunyaev76,Mushtukov+15}.
In this interpretation, all discovered PULXs have rather slow pulsations because pulsations with higher frequencies are washed out by the envelope. 

One way to address this controversy is to try to determine the mass accretion rate in an independent way and use it to calculate the spherization radius.
Since PULXs almost certainly have super-Eddington mass accretion rates, whether or not this material ends up onto the compact object, we expect that this large mass transfer produces visible effects on the binary system. 
Motivated by this, we obtained a long \nustar observation of the pulsar with the aim of detecting pulsations and, through the measurement of a possible orbital shrinking, constrain the total mass exchange in the system.
The observation was performed on UT 2016-04-15 -- 2016-04-19 (MJDs 57493.29 -- 57497.34).

\nustar does not resolve \Mtwo from another ULX, \Mone, only 5$\arcsec$ away.
This source is known to reach luminosities an order of magnitude higher than \Mtwo, and only \chandra is able to resolve the two ULXs.
Therefore, we undertook another program to monitor M82 monthly with \chandra with 25~ks per pointing with simultaneous \nustar  40~ks pointings.
This \chandra-\nustar program was designed to study the spectral evolution of the two sources, as well as to characterize the overall binary population in the M82 galaxy \citep[see][Brightman et al. in prep.]{brightman_60_2019}.

In this Paper, we present the timing analysis of the new data listed above and show that the pulsations are only detected in one new observation, despite the fact that, in some cases, we infer a high enough flux level from \Mtwo relative to \Mone from the \chandra observations for the pulsations to be detected.
Revisiting the work by B14, we show that the pulsed fraction evolves independently from the flux from the nearby \Mone, showing that this behavior is intrinsic to \Mtwo.
Thanks to the new detection, we are able to measure the orbital period of the system with greater precision, recover pulsations in one more old observation and detect a pulsar glitch.

In \sref{sec:datared} we describe the data reduction procedure.
In \sref{sec:search} we detail the pulsation searches performed and report on the new pulsations, the orbit and spin measurements.
We discuss the results in \sref{sec:discussion} and summarize them in \sref{sec:conclusions}.

\section{Data reduction}\label{sec:datared}
The observations considered in this work are detailed in Table~\ref{tab:allobsid}.
For \nustar, we used data processed with {\tt nupipeline} from HEASOFT v.6.25 \citep{centerheasarcHEAsoftUnifiedRelease2014}, with standard options. 
We barycentered old and new data in two independent ways: first, we used the FTOOL \texttt{barycorr} \citep{blackburnFTOOLSFITSData1995a} with the standard CALDB clock file; then, we tested a new temperature-driven model of the clock offsets, under development%
\footnote{The temperature-driven clock correction is now used to produce the official clock correction files with \nustar. 
The new clock correction is default since barycorr v2.2.
Bachetti, Markwardt et al. in prep}, 
and used PINT\footnote{www.github.com/nanograv/PINT} for the satellite orbit calculations.
The pulse period of \Mtwo is three orders of magnitude longer than the observed difference between the two barycentering methods.
For the orbit file, we used the attitude-orbit file produced by \texttt{nupipeline}%
\footnote{During the study of the temperature-driven model for the spacecraft clock, we realized that the orbit file distributed in the auxiliary data of \nustar observations, and recommended for use with barycorr, did not account properly for leap seconds.}. %
Running the pipeline using default values left some intervals of increased background activity (probably due to the South-Atlantic Anomaly), producing spikes in the light curve.
We verified that that the spikes correspond to increased activity in the entire field of view and not just from a single source; we removed the relevant intervals by modifying the good time intervals (GTI).
Following B14, we selected photons from a region of 70'' around \Mtwo. 
This region contains a large number of X-ray sources (e.g. B14, \citealt{brightman_spectral_2016}), with the total flux typically dominated by the combination of \Mone ($L_{\rm x,max} \sim 10^{41}$\,erg/s) and \Mtwo ($L_{\rm x,max} \sim 2\times 10^{40}$\,erg/s) \citep{brightman_60_2019}.
Given their $\sim$5'' separation, it is not possible to disentangle the contribution from these two sources to the \nustar data, so they must be considered together.

\begin{deluxetable*}{lccccccccc}
\tablecaption{Pulsation search results from all available observations of M82 X-2. \label{tab:allobsid}}
\tablecolumns{10}
\tablewidth{0pt}
\tablehead{
\colhead{ObsID} & \colhead{MJD} & \colhead{Date} & \colhead{Length\tablenotemark{a}} & \colhead{Exposure\tablenotemark{a}} & \colhead{Count rate\tablenotemark{b}} & \colhead{Meas. p.f.\tablenotemark{c}} & \colhead{Flux ratio\tablenotemark{d}} & \colhead{Corr. p.f.\tablenotemark{e}} & \colhead{$L_{39}$}\\
             &       &            &  (ks)  &     (ks) & (s$^{-1}$) & $\Delta F_{\rm X2}/F_{\rm tot}$(\%) &  $F_{\rm X2}/F_{\rm X1}$(\%) & $\Delta F_{\rm X2}/F_{\rm X2}$(\%) & 
}
\startdata
80002092002 & 56681 & 2014-01-23 & 123 & 66 & 0.9 & $<$5 & -- & -- & --\\
80002092004\tablenotemark{f} & 56683 & 2014-01-25 & 171 & 90 & 1.0 & $\phantom{<}$4 & -- & -- & --\\
80002092006 & 56686 & 2014-01-28 & 579 & 310 & 0.9 & $\phantom{<}$5 & -- & -- & --\\
80002092007 & 56692 & 2014-02-04 & 562 & 306 & 0.9 & $\phantom{<}$7 & 132$^*$ & $\phantom{<}$12 & 8.1$^*$\\
80002092008 & 56699 & 2014-02-10 & 62 & 34 & 1.0 & $\phantom{<}$7 & -- & -- & --\\
80002092009 & 56700 & 2014-02-11 & 213 & 115 & 0.9 & $\phantom{<}$9 & -- & -- & --\\
80002092011 & 56720 & 2014-03-03 & 201 & 111 & 0.6 & $\phantom{<}$3 & -- & -- & --\\
50002019002 & 57038 & 2015-01-15 & 56 & 31 & 0.7 & $<$6 & 40 & $<$27 & 5.7\\
50002019004 & 57042 & 2015-01-19 & 283 & 161 & 0.7 & $<$4 & 33$^*$ & $<$14 & 2.0$^*$\\
90101005002 & 57194 & 2015-06-20 & 56 & 37 & 2.3 & $<$3 & 32 & $<$17 & 19.8\\
80202020002 & 57414 & 2016-01-26 & 66 & 36 & 1.5 & $<$4 & 0$^*$ & N.A. & 0.0$^*$\\
80202020004 & 57442 & 2016-02-23 & 61 & 32 & 1.3 & $<$5 & 52 & $<$18 & 13.9\\
80202020006 & 57483 & 2016-04-05 & 54 & 31 & 0.8 & $<$6 & 44 & $<$26 & 6.7\\
30101045002 & 57493 & 2016-04-15 & 350 & 189 & 1.0 & $<$3 & -- & -- & --\\
80202020008 & 57503 & 2016-04-24 & 67 & 40 & 1.1 & $<$5 & 79 & $<$13 & 10.8\\
30202022002 & 57543 & 2016-06-03 & 60 & 39 & 1.0 & $<$5 & 0$^*$ & N.A. & 0.1$^*$\\
30202022004 & 57571 & 2016-07-01 & 68 & 47 & 1.7 & $<$4 & 33 & $<$17 & 15.1\\
30202022008 & 57599 & 2016-07-29 & 67 & 43 & 1.6 & $<$4 & 4$^*$ & N.A. & 1.2$^*$\\
30202022010 & 57619 & 2016-08-19 & 69 & 44 & 1.2 & $<$4 & 16$^*$ & $<$37 & 4.3$^*$\\
90201037002 & 57641 & 2016-09-10 & 94 & 80 & 1.2 & $\phantom{<}$3 & -- & -- & --\\
90202038002 & 57669 & 2016-10-07 & 71 & 45 & 0.9 & $<$5 & 5$^*$ & N.A. & 0.8$^*$\\
90202038004 & 57723 & 2016-11-30 & 68 & 43 & 0.8 & $<$5 & 0$^*$ & N.A. & 0.0$^*$\\
\hline
Det. puls. & & & & 1000 (58\%) & & & & &\\
Undet. & & & & 730 (42\%)  & & & & & \\
Low flux & & & & 207  & & &  & & \\
\hline
\enddata
\tablenotetext{a}{The length of the observation is the UT stop time minus the UT start time. The exposure is the actual on-source time not including occultation from the Earth and other ``bad'' intervals.}
\tablenotetext{b}{Count rates using FPMA + FPMB.}
\tablenotetext{c}{The pulsed fraction is referred to the total X-ray flux of M82, since M82 X-1 and X-2 are not separable in \nustar.}
\tablenotetext{d}{The flux ratio between \Mtwo and \Mone is estimated from \chandra data, either from the detailed and low-pileup spectral modeling of Brightman et al. in prep. when available, or from \citet{brightman_60_2019} otherwise (starred).
Dashes indicate that a \chandra dataset was not available.}
\tablenotetext{e}{A corrected estimate of the pulsed fraction, when a flux ratio is available. 
\textit{Not Allowed} (N.A.) indicates that the pulsed flux should be higher than the flux measured by \chandra.}
\tablenotetext{f}{Detected only after MJD 56683.5.}
\tablecomments{Intervals where \Mtwo is weaker than 10\% \Mone are not counted in the statistics of non-detections.}
\end{deluxetable*}

We did not redo the \chandra data reduction, and we used the data from \citet{brightman_60_2019} and Brightman et al. \textit{sub}.

\section{Search for pulsations and results}\label{sec:search}\label{sec:results}

We used a number of methods to detect new pulsations from \Mtwo, ranging from a focused search around the known frequency values and orbital parameters to a quasi-blind search (allowing for very large variations of the pulse frequency).

Data analysis was done through a set of custom python scripts based on HENDRICS\footnote{www.github.com/StingraySoftware/HENDRICS} \citep{bachetti_hendrics:_2018}, Stingray \footnote{www.github.com/StingraySoftware/stingray} \citep{huppenkothen-stingray_2019,huppenkothen_stingray:_2016} PINT \citep{luo_pint:_2019}, and PRESTO \citep{ransom_presto:_2011, ransom_new_2001}. 

In the 2-Ms long campaign performed in 2014, where the pulsar was first detected, pulsations did not have a constant r.m.s.
A large variation of pulsed fraction was reported by B14 and is also shown in \fref{fig:spindown}.
Orbital motion represented an additional difficulty.
In fact, in 2014 the pulsar was initially detected only after the pulsed fraction had increased well above the detection level, because orbital motion smears out the observed pulsed frequency, and the fainter signal was dominated by white noise.
Only after the first detection was made at high pulsed fraction, and the orbital parameters were determined, was the pulse found in some of the previous observations. 

Therefore, any search for pulsations in new observations had to account for at least two complications: weak pulsations, strong spin-up and orbital motion.
We used a two-tiered approach to this search: a deep search for pulsations around the expected spin period values and orbital parameters, and a more general search using multiple spin derivatives on 40-ks segments of data.

We report the details in the following subsections.

\subsection{Deep pulsation searches - first pass}
The first attempt consisted of a deep search of pulsations using the $Z^2_2$ statistics \citep{buccheri_search_1983}.
The $Z^2_2$ statistics was calculated from the functions in \texttt{stingray} and \texttt{HENDRICS}%
\footnote{
Rather than calculating the $Z^2_n$ statistics on the single events, this software pre-folds the data and then calculates the statistics using the phases of the profile with a weight given by the number of counts in each profile bin \citep{huppenkothen-stingray_2019}}.

We varied the frequency between the observed frequency in 2014 and the maximum frequency expected from a constant source spin-up of $5\times10^{-10}$\,Hz/s (more than double the maximum spin-up observed in 2014). 
The choice of the frequency interval was driven primarily by computing time and using an acceptable number of trial values.
The frequency step was 8 times finer than the standard (e.g. FFT) $\delta\nu=1/T$, with $T$ the observing time. 
This was to avoid the effects of spectral leakage on weak pulsations 
(e.g. see discussion on ``interbinning'' by \citealt{Ransom+02}). 
Not finding new detections above the standard $3-\sigma$ detection level accounting for the number of trials, we shifted the orbital phase by trial values spaced by $\sim$200\,s in order to account for a possibly imperfect orbital solution\footnote{200 s was chosen as it is approximately the error on $T_0$ that would produce a shift of $\sim$half a pulsar rotation over an orbit}. 
Again (and taking into account the increased number of trials), we did not find new significant pulsations with this strategy.

\subsection{Accelerated search}\label{sec:accel}
We used the PRESTO suite of pulsar search programs \citep{ransom_new_2001,Ransom+02} to run two different techniques of pulsation search, one Fourier-based (with the tool {\tt accelsearch}) and one epoch folding-based (with the tool {\tt prepfold}).

Following the strategy adopted in 2014, we split the light curves in segments of 30 ks, with sliding windows overlapping by a factor $\sim 0.5$.
This is motivated by the following: a longer light curve in principle allows for more signal to be accumulated in the periodograms, but orbital motion smears the signal of an accelerated search if the length is more than a certain fraction of the orbital period (the usual rule-of-thumb is 1/7 of $\Porb$).
This was clearly seen in the the 2014 campaign, where the maximum detectability with the accelerated methods in {\tt prepfold} was indeed obtained with segments of 30 ks, or about $1/7 \Porb$.

We binned the light curves to 1 ms, and produced binary floating point datasets in a format understandable by PRESTO using the \texttt{HENbinary} script in \texttt{HENDRICS}.

We first used the {\tt accelsearch} tool, that performs an accelerated search of pulsations based on the FFT and matched filters. 
We used the lowest detection limit ({\tt -sigma=1} on the command line), and specified that data were obtained by photons ({\tt -photon}). 
A number of ``candidates'' (possible signals above the threshold power value) were produced by this search and followed-up with \texttt{prepfold} to evaluate the significance using epoch folding.
All candidate periods from this search were very different from the reasonable interval of pulsations from \Mtwo.
We could also safely dismiss the possibility that these candidates were from other pulsars in the field of view, as the candidate pulsations always had very low significance and were not consistent between segments of data, as one would expect from pulsar candidates.
Some of these candidates are suspiciously close to beats of fundamental frequencies of the \nustar detector\footnote{e.g. a small dead time component that repeats with a frequency of $\sim890$\,Hz in datasets taken in CP mode, usually visible in very bright sources} or to the orbital data gaps.

We next tried to use the tool {\tt prepfold} to run an epoch folding-based search (i.e., suggesting candidate periods instead of letting \texttt{accelsearch} do so). 
\texttt{prepfold} automatically searches the $p-\pdot$ plane, including all reasonable $\pdot$ values expected from the source if the interval length is 30\,ks. 
Even if the maximum $\pdot$ in 2014 was high, many accreting pulsars have an average $\pdot$ that is much smaller than the instantaneous value found in single observations due to the alternation of spin-up and spin-down events that the accretion torque produces when accretion rate increases and decreases.
Therefore, we specified a starting value of the period around the mean value in 2014, and then used a number of starting values further and further from the mean value in both directions, randomly distributed in an interval larger than ten times the period variation expected if the maximum $\pdot$ in 2014 was constant until our new observations (which would result in spin periods of 1.31-1.372\,s). 
The random distribution allowed a certain amount of duplication (to test if a good candidate was consistently found with similar starting periods) and to avoid grid-related issues.
To evaluate a rough significance of pulsation candidates, we used the probability to find a given number of $\sigma$ from a search over 1000 realizations of white noise, using the same GTIs as the original observations. 
No significant ($>3\sigma$, using the above criterion) new pulsations were found.

\subsection{The ``Jerk'' search and a new detection in 2016}\label{sec:newdet}\label{sec:orbital}\label{sec:spindown}

Finally, we took advantage of the recently developed ``Jerk'' search technique in \texttt{accelsearch} \citep{andersen_fourier_2018}. 
This time, since this technique uses both the first and the second frequency derivatives, we used longer chunks of data, around 80\,ks (when the observation was long enough).
Thanks to the ``Jerk'' search, we found one new candidate pulsation in ObsID \texttt{90201037002}. 

\paragraph{New orbital solution}
\begin{figure*}
\fig{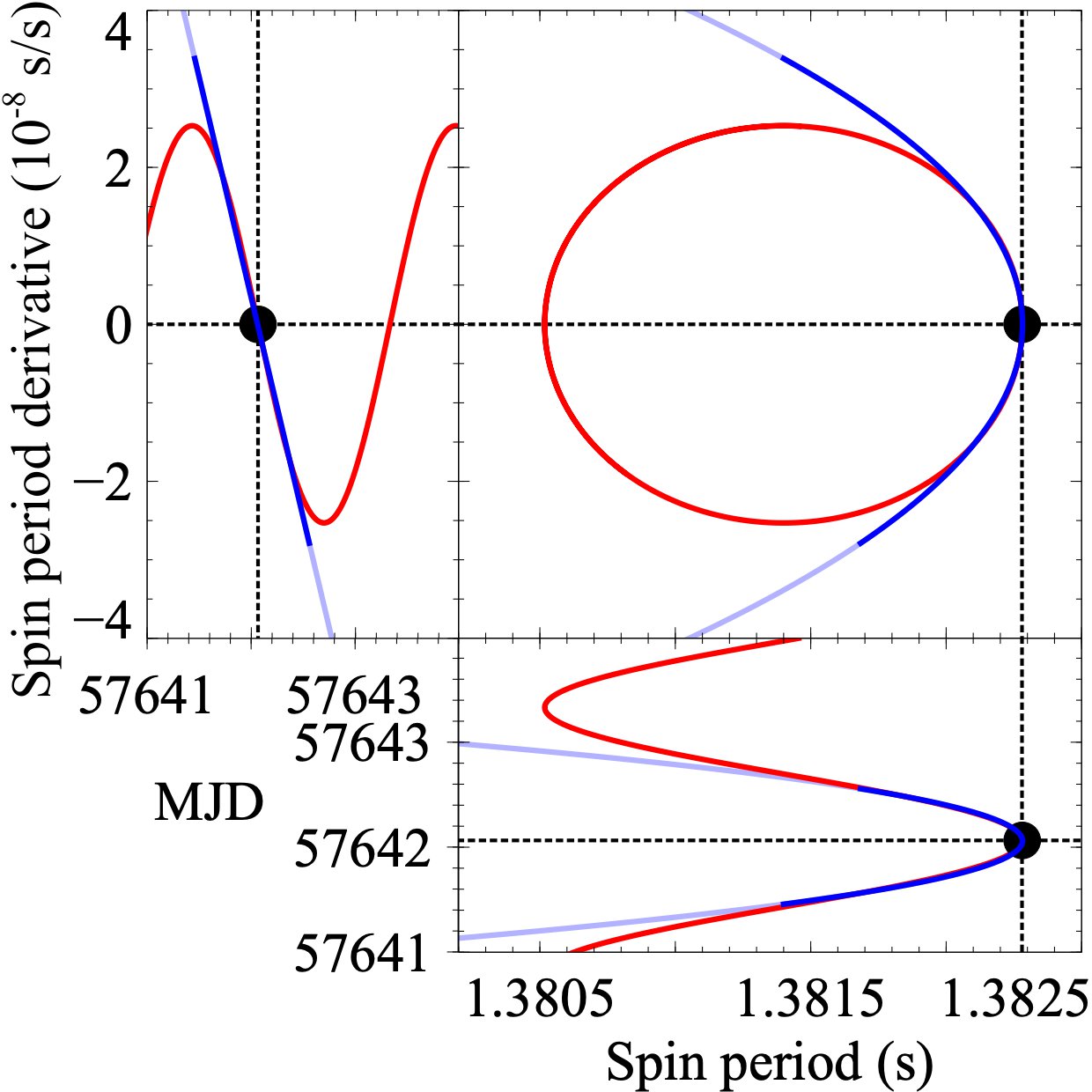}{0.381\textwidth}{a}
\fig{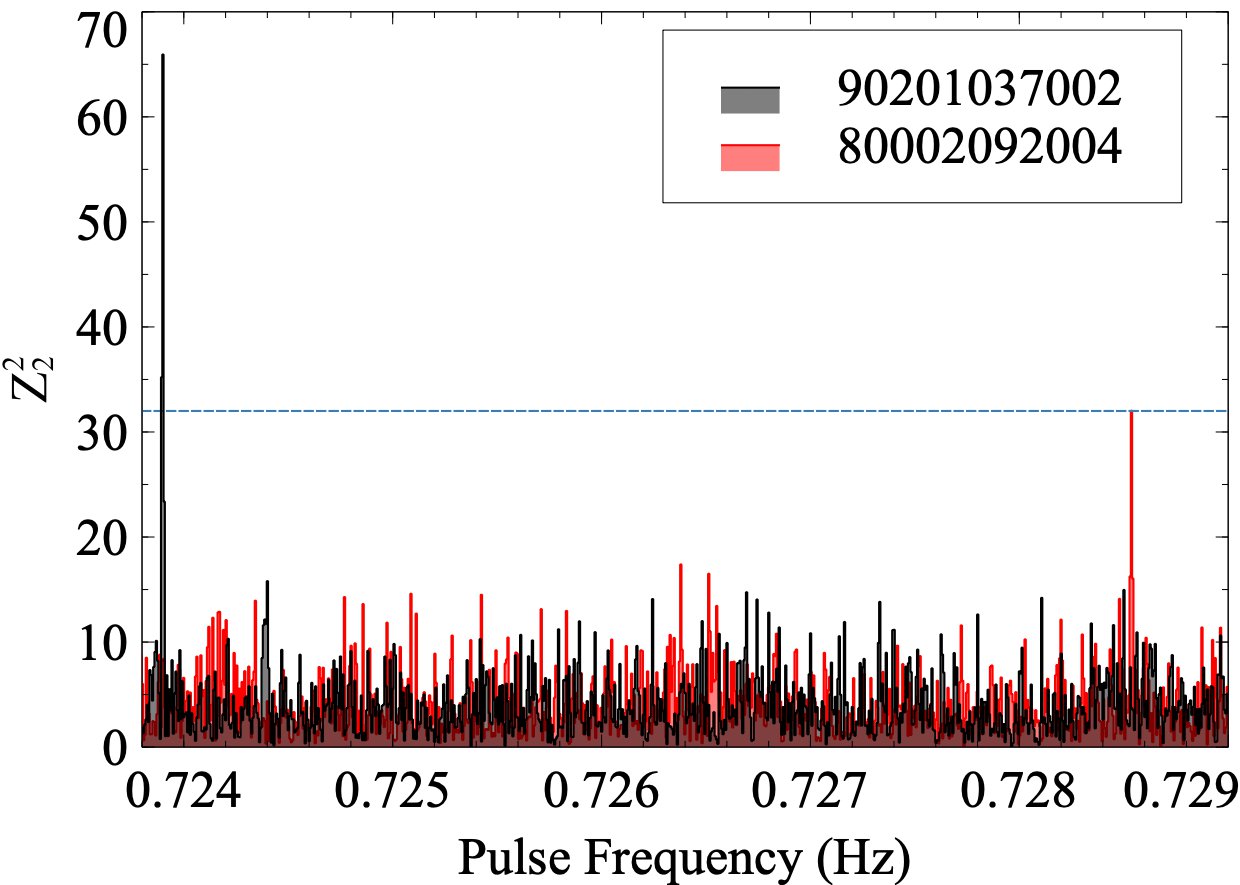}{0.525\textwidth}{b}
\caption{(a) The orbital phase can be measured from the spin solution if a measurement of at least two spin frequency derivatives ($\nudot$, $\ddot{\nu}$) is available.
Here we compare the best spin solution calculated by PRESTO (cyan, and blue in the time range covered by the data) and the expected $\nu$ and $\nudot$ change due to the orbital Doppler effect (red).
At this phase of the orbit, the constraint on the ascending node passage is very accurate.\label{fig:orbit}
(b) The new detections in ObsID 90201037002 and 80002092004, using the all-data orbital solution in \tref{tab:orbit}.
The horizontal line indicates the 3-$\sigma$ detection level for the given number of trial frequencies.\label{fig:newdet}}
\end{figure*}

\begin{figure}
\fig{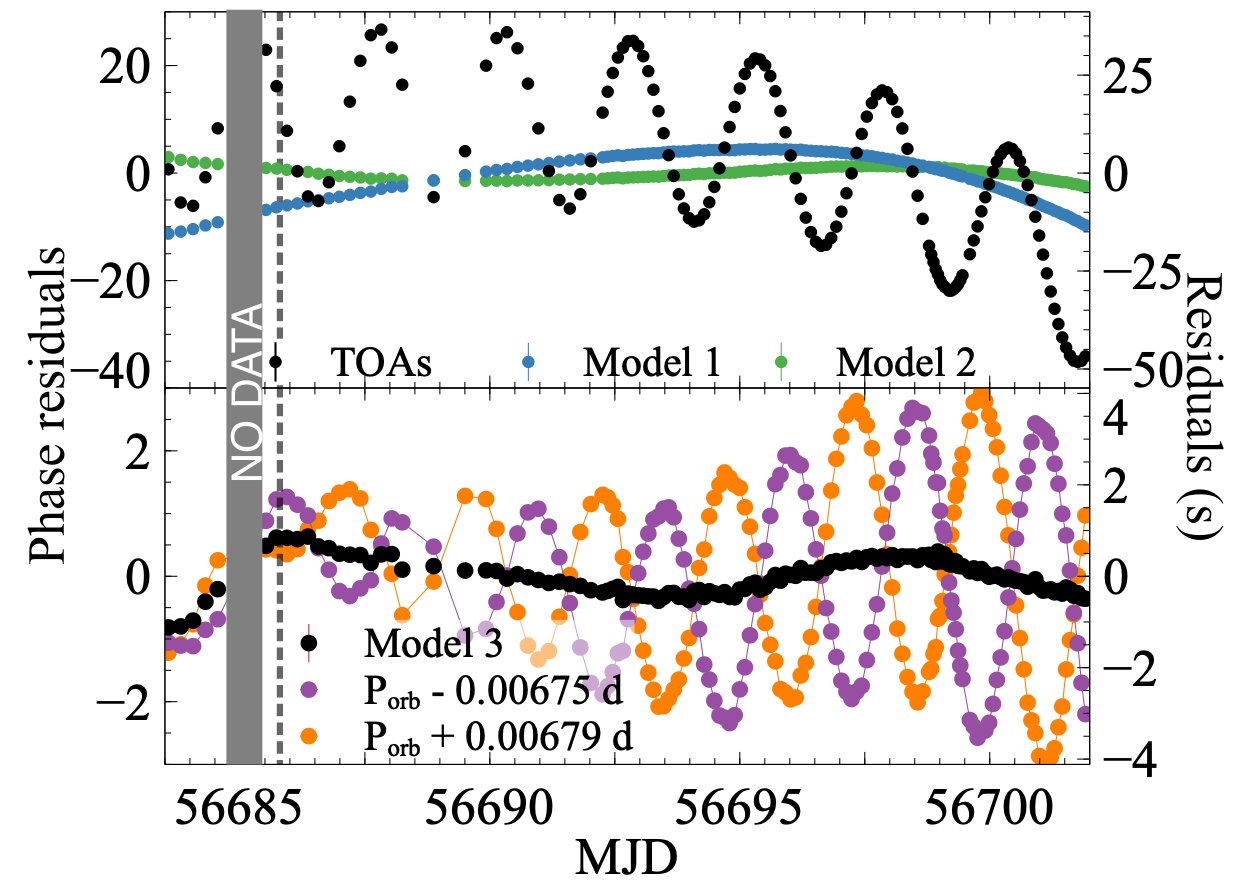}{\linewidth}{}
\caption{Top panel: residuals of the pulse times-of-arrival (TOAs) for the 2014 ObsIDs 80002092004-09 using the models in \tref{tab:spin}.
Black dots are the residuals from Model 1 if we ignore the orbital motion, and blue dots include the orbital motion.
Green dots add the $\nudot$ component (Model 2) instead.
Bottom panel: residuals of the pulsar TOAs using a timing model similar to Model 3, with two spin derivatives (F0, F1, F2), but changing the orbital period to accommodate one more (purple) or one less (orange) full orbit between ObsIDs \texttt{80002092011} and \texttt{90201037002}.
A very obvious modulation at the orbital period, with increasing amplitude, appears in the residuals.\label{fig:residuals}}
\end{figure}

We refined the candidate from the ``Jerk'' search with \texttt{prepfold}, and local values of the period and two period derivatives were measured with precision. 
They were consistent with those expected from a $\sim$1.37\,s pulsation, Doppler shifted by orbital motion with $\Porb\sim$2.52 d and $a\sin i/c\sim22$\,lt-sec, the orbital parameters known from B14. 
We refined the spin solution using the full observation (instead of a 80-ks segment) using \texttt{HENphaseogram}%
\footnote{The signal-to-noise in single intervals was not adequate to make more sophisticated timing using, e.g. \texttt{PINT} or \texttt{tempo2} at this step}%
.
These local spin derivatives give effects two orders of magnitude above the reasonable interval for the spin-up of the source, which is $|\nudot|\lesssim 2\times10^{-10}$\,Hz/s. 
Despite the observation covering less than 1/3 of an orbit, the measured spin derivatives were sufficient to put a constraint on the orbital phase (\fref{fig:orbit}). 
We can expect intrinsic (torque-driven) spin-up or spin-down whose magnitude is up to some $10^{-10} $\,s/s. 
The spin-up parameters and the orbital parameters are degenerate, and we include these considerations when calculating the improved error bar on the orbital period \Porb. 

Nonetheless, the 2.5-year lever arm between this new measurement and the original 2014 dataset allows for the error bar on \Porb to be significantly reduced (\tref{tab:orbit}).
With this rough orbital solution, we were able to fold the events in sub-intervals of the new observation and calculate the arrival times of the pulses with the \texttt{fftfit} method \citep{Taylor92} as implemented in \texttt{HENphaseogram}.
Then, we used \texttt{PINT} to refine the orbital and spin parameters, fitting for the frequency, the first derivative and the orbital period.
Given the small observation length, the spin derivative is not constrained by the fit, therefore we fixed it at 0.
Finally, we found a well-constrained solution and an improved value of \Porb. 
This new value is close to the one calculated from 2014 data, even if it is outside the range allowed by the quoted error bar on \Porb from the solution by B14. 
We will discuss the possible implications in \sref{sec:equilibrium}.

Later, we tested the possibility that an integer number of additional orbital periods could fit into the time range between ObsIDs \texttt{80002092011} and \texttt{90201037002}. 
An error on the period of $\sim$0.0001\,d, as implied by an additional or missing full orbit in this time range, would produce a very obvious distortion of the 2014 solution, and this possibility can be safely neglected (See \fref{fig:residuals}, bottom panel).

\paragraph{spin-down}
The most intriguing and direct consequence of this new detection is that, contrary to all other PULXs found up to now, the source showed an average spin \textit{down} (\fref{fig:spindown}) of $\nudot \sim -5.8\times10^{-11}$~Hz/s between 2014 and 2016.
The spin-down observed in this time range is likely driven by the negative torque of an accretion disk.
The alternating behavior between spin-up and spin-down suggests that the source is close to spin equilibrium.
An additional component could be dipole radiation from a strong magnetic field. 
We discuss the implications of this finding in \sref{sec:equilibrium}. 

\subsection{Deep pulsation searches with new orbital solution, and upper limits.}\label{sec:finaldeep}
Based on the improved orbital solution described in \sref{sec:orbital}, we ran a deeper search using the $Z^2_1$ (Rayleigh test) and $Z^2_2$ statistics \citep{buccheri_search_1983}. 
Given the unexpected spin-down, this new search included a much larger interval of frequencies but with a better constraint on the orbital phase.
This time, we also ran simulations to evaluate the pulsed fraction upper limit for the non-detections.
To do so, for each dataset, we maintained the GTIs and the total number of photons in each interval and randomized the time of the events inside each GTI. 
There is no significant broad-band noise at the frequency of the pulsar.
Again, we used the statistical functions in \texttt{stingray} and \texttt{HENDRICS}. 
As discussed above, this software allows the statistics to be calculated from the folded profiles instead of the single events as in the original Rayleigh test and \citet{buccheri_search_1983}.
To further speed up the calculation (in particular when calculating the larger number of realizations required by the upper limit calculations), we executed the total folding by folding $M$ equal-length sub-intervals of the observations, and aligning the folded sub-profiles differently for slightly different values of the frequency (similarly to the Fast Folding Algorithm, \citealt{Taylor+82}, and the technique used in \texttt{prepfold}). 
A bonus of this technique is that it is easy to shift the sub-intervals by integer values according to linear, quadratic, and higher-order laws and to measure rapid changes of trial spin frequencies.
The main rules-of-thumb to avoid dispersion of the signal in multiple bins are 1) that the number of bins in the profile $N$ is large enough that the pulse shape features are distinct in different bins and 2) that the number of profiles that are being shifted is at least twice the maximum shift of the sub-profiles.
We verified that calculating these statistical tests this way does not depart significantly from the expected statistics (More details in the Appendix and \fref{fig:detlev}).
Note that this new folding algorithm, as of August 13$^{\rm th}$, 2019, is made available in the development version of \texttt{HENDRICS}.

\begin{figure*}
\fig{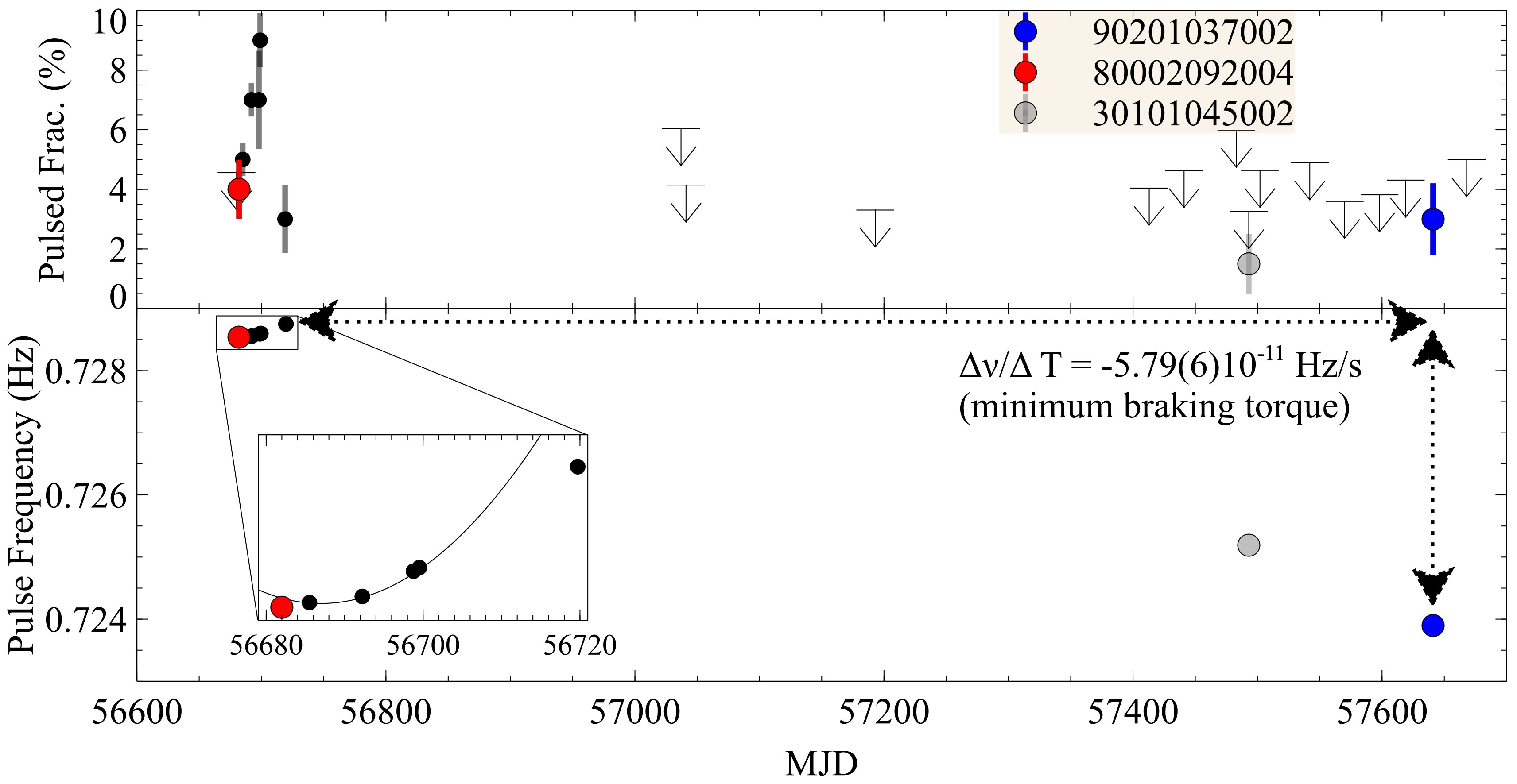}{\textwidth}{}
\caption{History of detections of pulsations from \Mtwo, showing the mean spin-down of the neutron star. 
The new detections are highlighted with larger markers (including the marginal one in grey).
The inset shows the timing solution from Model 3 in \tref{tab:spin}, applied to the data from ObsIDs \texttt{80002092004--09}.
Upper limits are indicated by arrows. Note that for the marginal detection in 30101045002 we plot both the marker and the upper limit.\label{fig:spindown}}
\end{figure*}

Using this method, we do not find  significant detections in the new observations.
As expected from the decrease of trial values and the better description of the data given by the orbital solution instead of a few spin derivatives, the detection in \texttt{90201037002} turns out to be highly significant (\fref{fig:newdet}).
We only find a hint of pulsations in ObsID \texttt{30101045002}, close to the predicted spin-down line and to the detection limit, that we plot in \fref{fig:spindown} with a grey point.

Surprisingly, despite not finding the pulsations in any more new observations, we did find a new significant detection in ObsID \texttt{80002092004}, the second observation of the 2014 campaign, where pulsations were not detected by B14. 
The reason why this happened is probably because the pulsation departs strongly from the smooth timing noise that was measured back then, shortly after the start of ObsID \texttt{80002092006}, as we will discuss in \sref{sec:glitch}.

We also repeated the search after filtering the events more aggressively by energy, only considering photons in the 8--30\,keV interval. 
While in ObsIDs \texttt{800020920XX} \Mtwo dominated the total flux and all emission between 3 and 30 keV was significantly pulsed, the pulsations in ObsID \texttt{90201037002} are only detectable above 8 keV (which is understandable by looking at the spectra reported by \citealt{brightman_spectral_2016} and following papers).
This new search, besides improving the marginal detection in ObsID \texttt{30101045002}, did not produce more detections.

\subsection{A glitch}\label{sec:glitch}

\begin{figure}
\fig{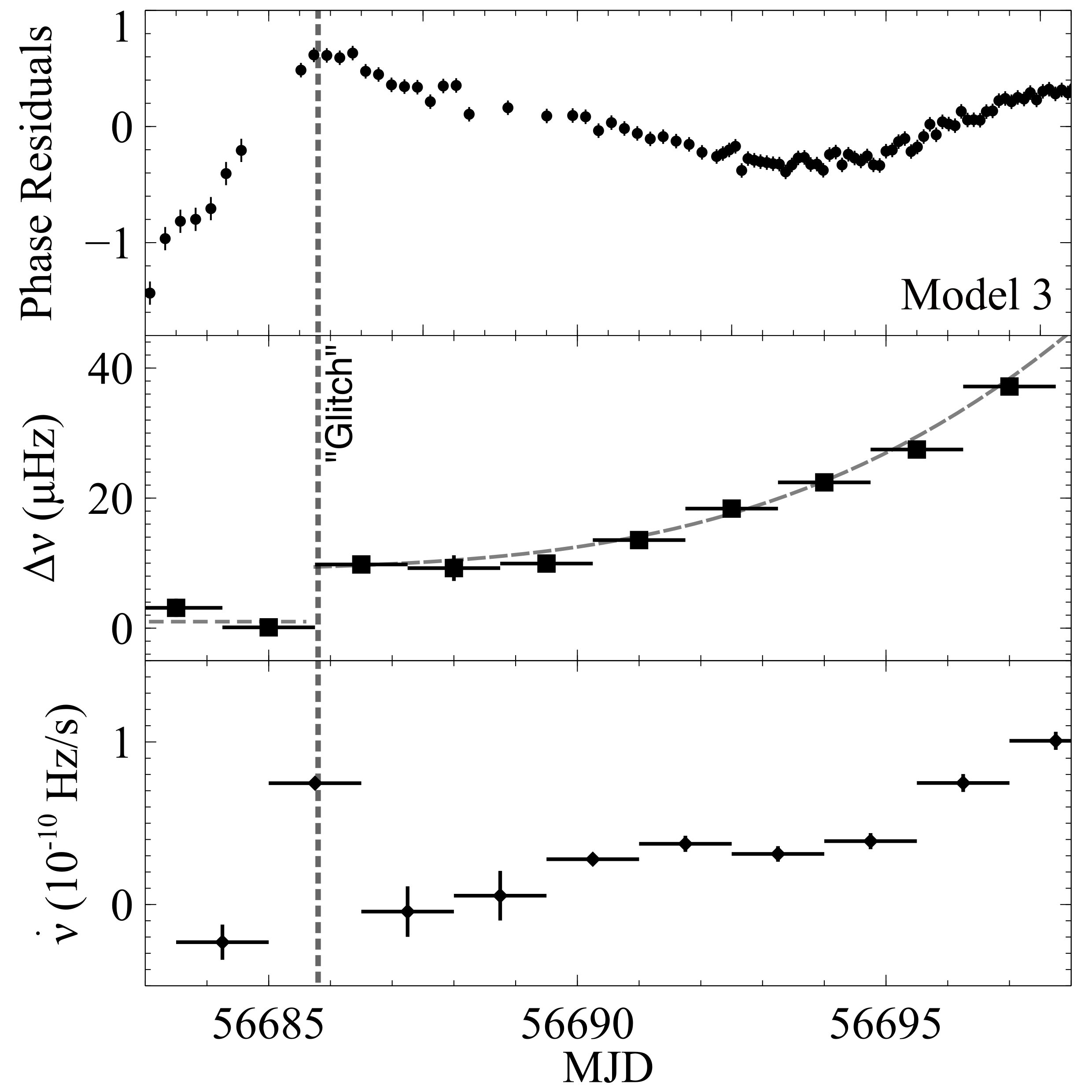}{\linewidth}{}
\caption{Top panel: TOA phase residuals after correcting the timing solution with Model 3 in \tref{tab:spin}.
Middle panel: local frequency variation, calculated with the \texttt{glitch} tool of \texttt{tempo2} using non-overlapping subsets of TOA residuals. 
The rapid change of frequency around MJD 56685.7 (vertical line), $\sim10 \mu$Hz, is clearly off with respect to the much smoother (but implying already quite extreme spin-up) frequency variation at later times.
Bottom panel: $\nudot$ calculated from the values of $\nu$ in the middle panel. 
The minimum average $\nudot$ necessary to produce the observed change of frequency between MJD 56685 and 56686.5 is $\sim8 \times10^{-11}$ Hz/s, very much above the local values of $\nudot$.
\label{fig:glitch}}
\end{figure}

Assuming the orbital solution given in \tref{tab:orbit}, we fit the data with timing models including an increasing number of spin derivatives (\tref{tab:spin}). 
After adding the F2 ($\ddot{\nu}$) term, besides additional long-term trends that might in principle be described by additional spin derivatives, an abrupt spin frequency change becomes apparent between ObsID \texttt{80002092004} and the start of \texttt{80002092006}.
Using the \texttt{glitch} plugin of TEMPO2, we calculate two local spin derivatives (F0 and F1) from the pulse times of arrival.
\fref{fig:glitch} shows the result: we find that the timing solutions need either a sudden frequency jump or a very strong frequency derivative at around the same time.
The measured frequency change is too sharp to be described by a small number of additional spin derivatives. 
It amounts to around $10^{-5}$ Hz, corresponding to $\Delta\nu/\nu\approx 1.3(4) 10^{-5}$ (\tref{tab:spin}).
The minimum $\nudot$ necessary to produce such a large frequency increase on timescales of ~1 day is $\sim 10^{-10}$ Hz/s.
Based on simple arguments from accretion theory, such a strong frequency derivative would need a significant change of mass accretion rate and, presumably, of luminosity, that we do not observe.
We therefore discuss from now on the simplest explanation, i.e. that the neutron star has undergone a ``glitch'', a sudden change of spin velocity.

The measured frequency change of $\Delta\nu/\nu\gtrsim 10^{-5}$ is a very large value, but not unprecedented, for a pulsar glitch. 
Typical glitch magnitudes in radio pulsars are more than an order of magnitude (often many orders of magnitude) smaller, but those observed in magnetars and accreting pulsars can reach those values; see \citealt{manchester_pulsar_2018} for a review, and \sref{sec:glitchdisc} for a discussion.

To exclude instrumental effects, we checked the details of the clock correction file during this time range. 
We compared the results of the standard clock correction pipeline with a new clock correction based on an improved thermal model of the spacecraft oscillator (Bachetti et al. in prep.), finding a very good agreement (well inside the error bars).
Bad clock offset measurements from the ground stations could in principle produce changes in the measured pulsed frequency, but we verified that the scatter of the offset measurements from the Malindi station is $\leq$100 $\mu $sec and we excluded all measurements from other, less reliable, ground stations, with no significant changes in the results.

\begin{deluxetable}{lccc}
\tablecaption{Updated Orbital parameters of M82 X-2, compared to the B14 fit and to a new fit of the 2014-only data.
The quoted uncertainties represent approximate symmetrical 1-$\sigma$ confidence intervals, as calculated by \texttt{PINT}.
Values and error bars on $a\sin i$ and $T_0$ are fixed to the 2014-only value.
The 2014-only quoted error bar is based on the orbital phase measured in ObsID \texttt{80002092011}, with 40-d lever arm with respect to $T_0=$MJD 56682.0661, and including the local first derivative in the fit.
The full 2014--2016 solution is based on the orbital phase measured in \fref{fig:orbit} and refined with PINT, fixing the first derivative to zero (see text).
\label{tab:orbit}}
\tablecolumns{4} 
\tablehead{
\colhead{Parameter} & \colhead{B14} & \colhead{2014-only fit} & \colhead{All data}}
\startdata
$T_0$ (MJD) &  56694.7327(1) & 56682.0661(3) &  56682.0661(3)\\
$a \sin i$ (lt-s) & 22.225(4) & 22.215(5) & 22.215(5)\\
$P_{\rm orb}$ (d) & 2.53260(5) & 2.53287(5) & 2.532948(4)\\
\hline
\enddata
\end{deluxetable}

\begin{deluxetable*}{lcccc}
\tablecaption{Spin parameters for the observations in 2014 using different models. Orbital parameters were fixed to the values in \tref{tab:orbit}
\label{tab:spin}. The reference epoch for all models is MJD 56683.
Note that, even if the F1 term is negative suggesting spin-down, the effect of the second spin derivative F2 is dominant and gives spin \textit{up}.}
\tablehead{Parameter & \colhead{Model 1} &  \colhead{Model 2} & \colhead{Model 3} & \colhead{Model 4}
}
\startdata
F0 (Hz)         & 0.728566160(7) & 0.72852171(3) & 0.72855082(8) & 0.72854408(11) \\   
F1 ($10^{-11}$\,Hz/s)       & 0	             & 4.758(3)  & -2.98(2)  & -5.29(3) \\   
F2 ($10^{-17}$\,Hz/s$^2$)   & 0	             & 0         & 8.66(2)   & 10.93(3) \\   
GL\_F0 (Hz)      & 0	             & 0             & 0             & 1.75(2)$\times 10^{-5}$\\   
GL\_EP (MJD)  & --	             & --             & --             & 56685.8 (fixed)\\   
r.m.s. ($\mu s$) & 136	         & 34            & 9.2           & 4.3\\   
$\chi^2$ ($\mu s^2$) & 2698012   & 165709        & 12251         & 2704\\
\enddata
\end{deluxetable*}

\section{Discussion}\label{sec:discussion}
The strong spin-down observed between 2014 and 2016, the dramatic spin-up and pulsed fraction variation during the 2014 observation, and the non-detection of pulsations in subsequent observations even with comparable fluxes of \Mtwo measured by \chandra, can shed light on \Mtwo and the ULX phenomenon on the whole.
The glitch revealed in the 2014 observation also adds to the list of peculiar phenomena observed in ULXs, following the detection of anti-glitches in NGC 300 X-1 \citep{ray_anti-glitches_2018}.

In this Section, we discuss how these new data can shed light on the nature of \Mtwo and the ULX phenomenon.

\subsection{Caution: luminosity and accretion rate in \Mtwo}\label{sec:flux}

Most of the interpretation in the following subsections, in particular for what concerns the transient pulsations and the spin-up, could be done by simply assuming that the luminosity is a simple function of mass accretion rate. 

However, this needs a few words of caution.
An example of why this simple assumption might be flawed can be found in \citet{vasilopoulosNGC300ULX12019}: the authors find that, despite a large drop of luminosity of the source NGC 300 ULX1, the pulsar keeps spinning up with the same rate as when the luminosity is high. 
This is probably due to the fact that the luminosity drop is given by occulting material, like a super-Eddington wind, and not an actual change of accretion rate.

If we take for granted that the the drop of flux of \Mtwo corresponds to a change of accretion rate, the finding by \citet{brightman_60_2019} that this flux change is periodic would imply that the mass accretion rate changes periodically, switching accretion on and off every $\sim60$\,days.
For the physical properties of this system, however, this is problematic.
While it would be easy to produce a a periodically-changing mass transfer rate if the period corresponded to the orbital period, this is far from obvious for a super-orbital periodicity.
Assuming Roche Lobe overflow, possible mechanisms to increase periodically the pressure at the L1 Lagrangian point might involve additional orbiting objects, like the so-called Kozai mechanism \citep{lidov_evolution_1962,kozai_secular_1962} or tidally excited oscillations \citep[e.g.][]{fuller_heartbeat_2017}. 
Note that a very small periodic change of orbital eccentricity, well below the quoted upper limit from B14, would be sufficient to produce substantial changes of accretion rate \citep[e.g.][]{hut_effects_1984,maccarone_fading_2010}.
In the same way, a stellar pulsation could produce a periodic increase of the mass accretion rate.
However, we can exclude an evolved star due to the the small semimajor axis of the orbit \citep{fragos_formation_2015,heida_searching_2019} and the likely range of companion masses derived from timing (5--20 $M_{\odot}$, B14).
For this range of masses, we expect the companion to be a main sequence O-type star, and pulsations reported for these systems (e.g. from $\beta$-Cepheids or slowly-pulsating Be stars) are on shorter timescales. \citep[see][]{samus_general_2017}.
Stars that can in principle have stable pulsations in the 60-d range are usually either evolved lower-mass stars or luminous blue variables (LBVs) that are thought to be much more massive than the likely 5--20\,$M_{\odot}$ range for \Mtwo \citep[see, e.g.][]{conroy_complete_2018,jiang_three_2018}. 
Note also that \citet{heida_searching_2019} rule out LBVs in the error circle of \Mtwo.

The point of these arguments is that historic luminosity changes in \Mtwo might be dominated by some occulting material, such as a precessing disk, and not by actual mass accretion rate changes. 
The variability of \Mone complicates the investigation further, and undermines a precise interpretation of the pulsar behavior based on luminosity.

Future studies will address this problem. 
For now we keep both doors open and discuss the consequences of both hypotheses.

\subsection{Pulsed fraction evolution and non-detections}

The pulsed fraction of \Mtwo changed dramatically in 2014, and in the new observations it often showed no detectable pulsations. 
In \tref{tab:allobsid} we report the estimated pulsed fraction and/or upper limits on pulsations, plus the luminosity from the PULX \Mtwo and the flux ratio between \Mtwo and \Mone when \chandra observations are available.
In the latter cases, we also correct the upper limit of the pulsed fraction.
The detectability of pulsations is largely driven by the intensity of the source $I$, the background $B$ and the pulsed fraction or equivalently the r.m.s. $r$, and only weakly by the observing time $T$. 
\citet{Lewin+88} show that the significance of pulsations goes roughly as $I^2/(I+B)r^2T^{1/2}$.
Looking into detail at the table, we see that 2014 observations were executed in a very favorable situation, with \Mtwo brighter than \Mone during ObsID 80002092007, a high (even if not at the highest values) flux of \Mtwo and a very long observing time.
The pulsed fraction change in 2014 is positively correlated with a flux increase of \Mtwo (see below), and we could have expected even higher values in these later observations.

\begin{figure}
\includegraphics[width=\linewidth]{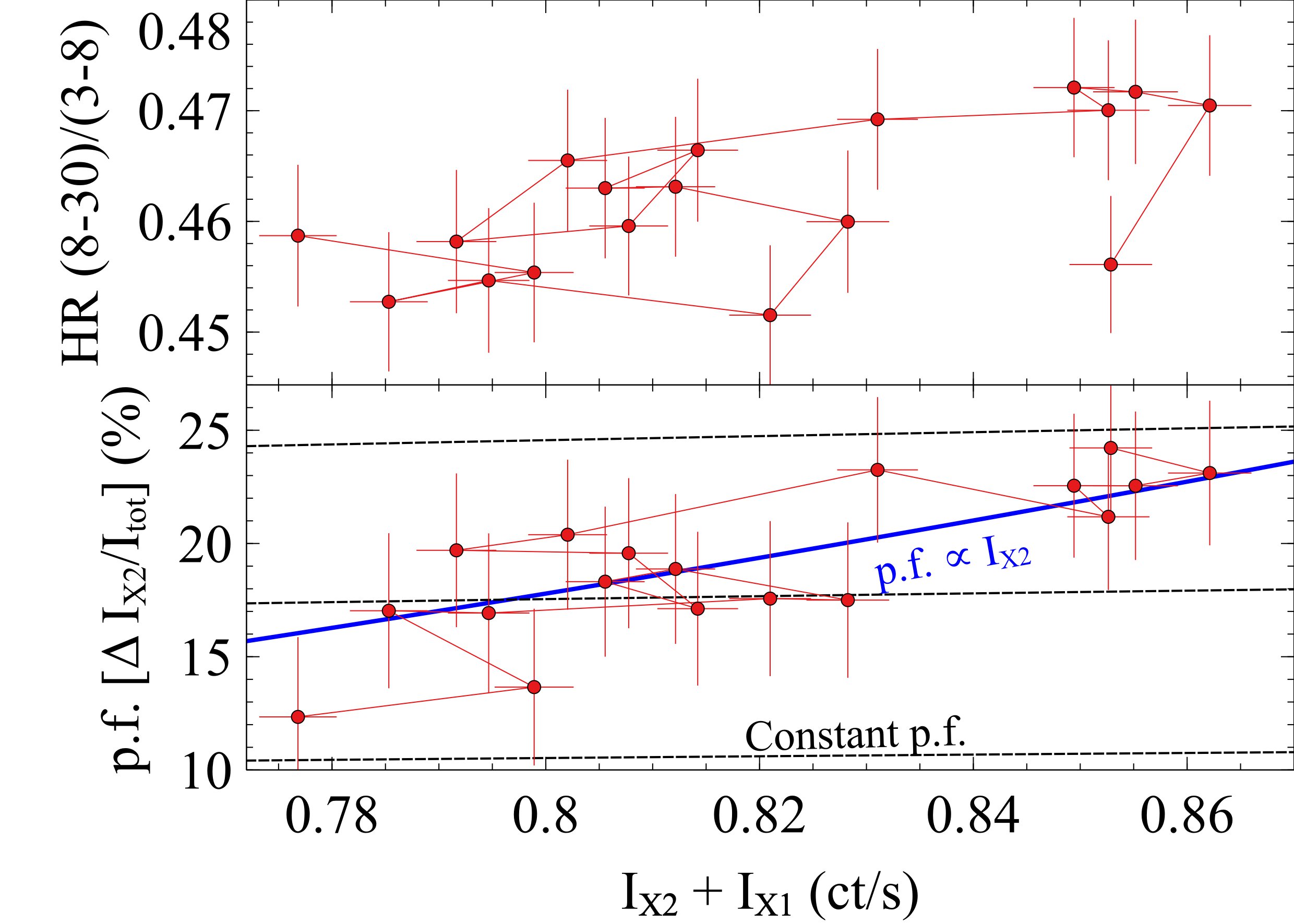}
    \caption{Each point represents 80\,ks of data during the 2014 campaign. 
    The pulsed fraction increases with total flux. This implies that the total flux increase is driven by \Mtwo, not \Mone, and allows us to assume intensity as a proxy for accretion rate for the 2014 data.
    Also, the increase of total pulsed fraction (pulsed flux over the total flux of \Mone and \Mtwo) is too steep to be justified by just the flux evolution of \Mtwo over \Mone (the dashed lines show the expected increase in this case).
    There seems to be an approximately linear increase of pulsed fraction with \Mtwo's intensity (blue line).
    Hardness increases slightly with intensity, as expected from an increase of \Mtwo, which is generally harder \citep{brightman_spectral_2016}.
    \label{fig:pulsestats}}
\end{figure}

\fref{fig:pulsestats} shows that the increase of pulsed fraction is associated with an \textit{increase} of flux and a stable hardness ratio. 
This tells us that the pulsed fraction is increasing because \Mtwo is brightening and not, e.g., because of a weakening of \Mone. 
Therefore, it is safe to associate the change of luminosity with a change of accretion rate, and to intepret the increasing torque implied by the strong second derivative $\ddot{\nu}$ in these terms as we do in \sref{sec:equilibrium}.

On the contrary, a few observations in the following years were clearly in unfavorable conditions for pulsation detection. 
In ObsIDs 80202020002, 30202022002, 90202038004 the flux of \Mtwo was extremely low.
In 2015--2016, \Mone underwent some flaring events, and pulsations would have been impossible to detect in ObsIDs 30202022008 and 90202038002 despite the source luminosity being $\gtrsim 10^{39}$\,erg~s$^{-1}$.
The low pulsed fraction measured in ObsID 90201037002 probably implies that \Mone was strong with respect to \Mtwo, but not strong enough to completely drown the pulsations. 
On the other hand, the detection of pulsations tells us that \Mtwo was stronger than in the two observations performed approximately two weeks before and after, 30202022010 and 90202038002, despite the overall \nustar count rate being similar\footnote{Note that change of flux from \Mtwo on these time scales is compatible with the super-orbital period of 60 days shown by \citet{brightman_60_2019}.}.

In ObsIDs 90101005002, 80202020004 and 80202020008 the flux of \Mtwo was higher than in 2014.
The flux of \Mone was also higher in these observations, and we would not expect the same significance as 2014 but still, we did not find credible candidate pulsations, even marginal, in these ObsIDs, even if the pulsed fraction upper limit is compatible with the values measured in 2014.
Also, 80002092002 and the start of 80002092004 show no detectable pulsations. This might be due to the onset of accretion, with the higher flux in 80002092002 related mostly to \Mone.
\Mtwo is not the only ULX showing intermittent pulsations.
Recently, \citet{sathyaprakashDiscoveryWeakCoherent2019} found intermittent pulsations in NGC 1313 X-2. 
ULXs are peculiar objects, probably Roche-lobe filling like low-mass X-ray binaries (that have much lower companion masses) but with companion masses more similar to high-mass X-ray binaries (that mostly accrete via wind)\footnote{For completeness, we mention that \citet{mellahWindRocheLobe2019} proposed an intermediate Roche-wind accretion model, initially proposed by \citet{mohamedWindRocheLobeOverflow2007}, for ULXs}.

Therefore, it is useful to look for transient pulsations in both kinds of systems, and see what processes might be at work.
One notable example is Aql X-1, an accreting millisecond pulsar in a low-mass X-ray binary, probably with a relatively low magnetic field ($10^{9}$G) and sub-Eddington.
\citet{Casella+08} reported pulsations lasting only $\sim150$\,s over several megaseconds of \rxte observations. 
Another accreting millisecond pulsar, HETE J1900.1-2455, showed transient pulsations during an outburst, with changes of pulsed fraction associated with thermonuclear bursts \citep{galloway_intermittent_2006}.
A0538-66, an HMXB, showed pulsations in 1982 \citep{skinner_discovery_1982} and then never since \citep[see][for a review]{kretschmar_xmm-newton_2004}.
Recently, \citet{brumback_discovery_2018} and \citet{pike_observing_2019} reported on transient pulsations from the X-ray binaries LMC X-4 and SMC X-1.
In the first case, a significant change of pulse properties (r.m.s., spin derivative) was associated with the precursor of a large burst-like flux increase, which is compatible with an increase of the accretion rate at the surface (e.g., the disk overcoming the centrifugal barrier of the magnetized neutron star).
In the second, there was, instead, evidence of a change of absorption between the non-pulsed and the pulsed intervals, suggesting the presence of occulting material.

A precessing structure that occults the central X-ray source could be the explanation for the periodic modulation of the flux from \Mtwo \citep{middletonDiagnosingAccretionFlow2015,brightman_60_2019}.
A strong disk wind, as observed in several ULXs \citep[e.g.][]{pinto_ultrafast_2016,Fabrika+15,kosec_evidence_2018}, would follow a precessing disk \citep[e.g.][]{pashamCan62Day2013,middletonLenseThirringTimingaccretionPlane2019} and block the view of the central flux from the compact object.
For example, \citet{dauser_modelling_2017} model long super-orbital periods observed in ULXs as the precession from a disk launching a conical wind.
Precession is observed in an archetipal super-critical source, SS433, and various ULXs \citep{begelman_nature_2006}.
It is expected to produce a periodic hardening of ULX spectra \citep{middleton_spectral-timing_2015}.

Using the best \chandra data available, Brightman et al. in prep. do not find a significant spectral evolution of \Mtwo, and in particular, no evidence of large variations of $N_H$. 
The spectrum is often consistent with the pulsed spectrum reported by \citet{brightman_spectral_2016}.
However, the $N_H$ is quite high and the spectrum of \Mtwo is hard, with the Galactic X-ray emission dominating the lowest energies below the \nustar band.
It is possible that changes of $N_H$ go unnoticed or confused with other degenerate parameters.
Also, a highly ionized wind would reduce the flux by scattering rather than the photoelectric effect, reducing the measured $N_H$.

An additional explanation might involve the precession of the pulsar, with the pulse beam moving away from the observer.
In principle, this might be detected through changes of the pulsar spectrum.
Given the relatively low count rate of the source, and the large source confusion (with \Mone at 5$''$), this is not testable on \nustar data and difficult even from existing \chandra spectral data that are sparse and often affected by pile-up, which affects the higher energies where the pulsed component is more significant.
In principle, the large spin-down discussed in the next Section might indicate a change of the accretion rate that could indeed alternate between accretion and no accretion, with the total emission dominated by a disk close to spin equilibrium.
With the available data, it is not possible to get a definitive answer.

\subsection{Spin behavior: not a slow rotator}\label{sec:equilibrium}

\begin{figure}
\includegraphics[width=\linewidth]{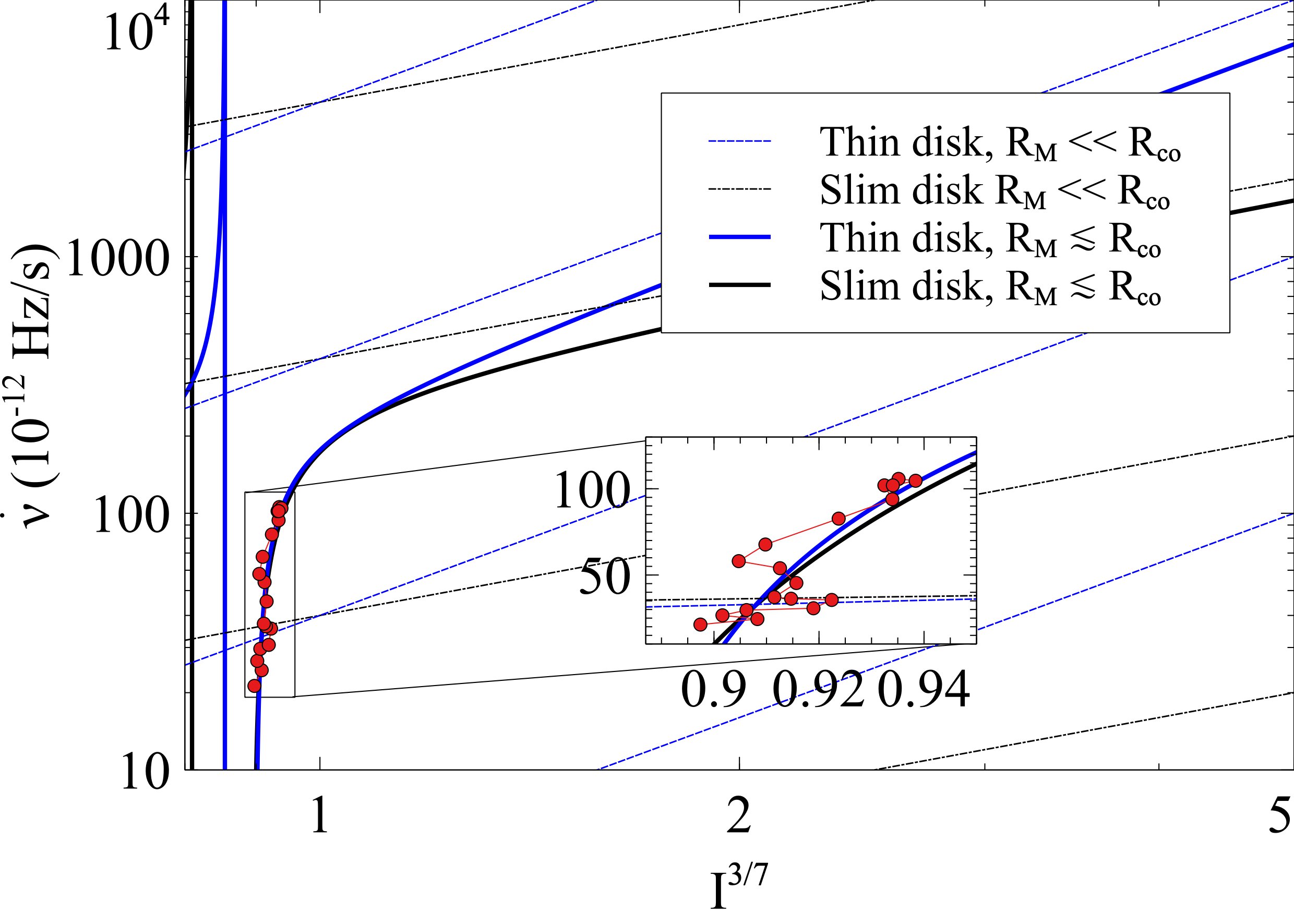}
    \caption{Using the same data as \fref{fig:pulsestats}, we compare the measured values of intensity and spin-up with the predictions of torque theories in the case of a \textit{slow} rotator, where $\rin << \rco$. 
    Far from spin equilibrium, we would expect the frequency derivative to be proportional to $\Mdot^{6/7}\propto L_{37}^{6/7}$ \citep{FrankKingRaine}. 
    For slim disks, $L\propto\Mdot^2$ \citep{kingAccretionRatesBeaming2008,kingPulsingULXsTip2017}. The data are incompatible with the slow rotator hypothesis, suggesting that $\rin \sim \rco$. 
    Indeed, using \eqref{eq:fastness} with $\fastness\sim1$, the curve can be made to agree with the data.
    \label{fig:i37}}
\end{figure}

The absence of pulsations for a high fraction of time, and the secular spin-down observed between 2014 and 2016, despite luminosity trends consistent with past observations, suggest that the source might be close to spin equilibrium, with large variations of spin-up or spin-down produced by relatively small changes of the magnetospheric radius $R_M$ around the corotation radius \rco.

Using the formulae by \citet{GL79b}, and assuming constant $\dot{M}$, one can show that the spin-up of a pulsar should follow a relation 
\begin{equation}\label{eq:pdot_over_p}
    -\dfrac{\pdot}{p} \propto p\,n(\fastness)
\end{equation}
where $p$ is the pulse period, $\pdot$ its first derivative, $n$ is the torque from the disk on the magnetic field lines and $\fastness=(R_M/R_{\rm co})^{3/2}$ the \textit{fastness parameter}. 
\citet{wang_torque_1995} derives the following expression for the torque:
\begin{equation}\label{eq:fastness}
    n(\fastness)\approx\frac{7/6 - (4/3)\,\fastness + (1/9)\,\fastness^2}{1- \fastness}.
\end{equation}

If the pulsar is a \textit{slow} rotator, with $\fastness<< 1$ and so very far from spin equilibrium, this relation reduces to a constant, $n(\fastness)\approx7/6$, and \eqref{eq:pdot_over_p} gives
\begin{equation}
    -\dfrac{\pdot}{p} \propto p
\end{equation}
This relation holds, for example, in NGC 300 ULX-1 \citep{vasilopoulos_ngc_2018,vasilopoulosNGC300ULX12019}.

However, the spin-up shown by \Mtwo during the 2014 campaign does not follow this relationship at all (\fref{fig:i37}).
If we consider a change of $\dot{M}$ proportional to the change of luminosity of the source, we expect a relation of the kind \citep[see][]{GL79b}
\begin{equation}
    -\dfrac{\pdot}{p^2} = \nudot \propto L_{37}^{6/7}\,n(\fastness),
\end{equation}
while in the ``slim disk'' case with beaming \citep[][]{kingAccretionRatesBeaming2008,kingMassesBeamingEddington2009}, $\dot{M}\propto L^{3/7}$ and
\begin{equation}
    -\dfrac{\pdot}{p^2} = \nudot \propto L_{37}^{3/7}\,n(\fastness).
\end{equation}
Again, \fref{fig:i37} shows clearly for both models that the spin behavior of \Mtwo in 2014 is inconsistent with $\fastness<<1$, while it can in principle be reproduced if we assume $\fastness\sim1$.
Therefore, the basic assumption of most concurring models on \Mtwo \citep[e.g.][]{Bachetti+14,Dallosso+15,tsygankovPropellerEffectAction2016,kingPulsingULXsTip2017}
Here we calculate spin derivatives from a robust interpolation\footnote{We use the method by \citet{savitzky_smoothing_1964}, as implemented in \texttt{scipy.signal}} of order 2 of single spin frequency measurements during the campaign.

The source might be undergoing some sort of spin reversal over time. 
This would be similar to what reported for LMC X-4 by \citet{molkov_near-periodical_2017}, where an alternating spin-up and down behavior was observed over many decades.
In that case, one of the possible explanations was a \textit{Recycling magnetosphere model} \citep{perna_spin-up/spin-down_2006} where accretion is only possible at certain spin phases.


Independently from accretion torque, the pulsar might be spinning down because of its own dipole radiation.
According to the classic magnetic dipole radiation formula\footnote{See any textbook on pulsars, e.g. \url{https://www.cv.nrao.edu/course/astr534/Pulsars.html}}, if the spin-down of the pulsar is given by dipole radiation, we can estimate the magnetic field as:
\begin{equation}
  B > \left(\frac{3 c^3 I}{8\pi^2 R^6} p\pdot\right)^{1/2} \approx 3.2\times 10^{19}\left(\frac{p\pdot}{\rm s}\right)^{1/2} G
\end{equation}
where $I$ is the moment of inertia of the neutron star and $R$ its radius.
Let us assume a fine tuned equilibrium from the accretion torque, with no net spin-up or down from accretion and the spin-down dominated by dipole emission. 
Substituting the spin period and the spin-down period derivative in the formula above, and using standard estimates for the radius and moment of inertia of the neutron star, we get a lower limit on the magnetic field of $\sim 3\times10^{14}$\,G.
This would be in line with the estimate made by \citet{Tsygankov+15}, interpreting the high- and low-luminosity states of the pulsar as alternating accretion-propeller regimes and assuming that there is no mass loss before accretion onto the pulsar.
Future robust measurements of the total mass exchange will give a better estimate of the position of the spherization radius, and will help to test the predictions of these and other competing models.

\subsection{Glitch: strong spin-down before 2014?}\label{sec:glitchdisc}

There are very few glitches reported from accreting neutron stars.
KS 1947+300 showed a large glitch of $\Delta\nu/\nu\approx 3.5\times10^{-5}$ in \rxte observations in the early 2000s \citep{galloway_frequency_2004}, comparable with $\Delta\nu/\nu\approx 10^{-5}$ reported in \sref{sec:glitch}. 
Similarly, \citet{serimDiscoveryGlitchAccretionpowered2017} report on a strong glitch from the slow accreting pulsar SXP1062, with $\Delta\nu/\nu\approx 10^{-3}$.
Besides SXP 1062 and KS 1947+300, glitches with $\Delta\nu/\nu\gtrsim 10^{-5}$ have only been reported in magnetars (e.g. \citealt{dib_rossi_2009,dib_16_2014,archibald_magnetar-like_2016}).
Some sources, and notably the ultraluminous pulsar NGC 300 ULX-1 \citep{ray_anti-glitches_2018}, show \textit{anti-}glitches instead -- sudden slow-\textit{downs} of the pulsar.
A non-magnetar glitch is believed to be due to the independent rotation of the NS core with respect to the rest of the NS. 
A neutron star contains superfluid material in the core and in the inner crust, and the different regions of the star interior are allowed to rotate with different velocities.
If the crust slows down over time, e.g. due to dipole radiation as in radio pulsars, the spin rates of the crust and the core differ more and more over time.
It is believed that when the differential rotation reaches some critical value, these regions connect, angular momentum is transferred between the crust and the core, and the star suddenly starts spinning as a whole.
If the crust was slowing down over time, the connection with the core will produce a sudden acceleration.
The opposite is believed to have happened with the anti-glitch observed in NGC 300 ULX-1 \citep{ray_anti-glitches_2018}.
In that case, the star has been observed to spin-up over time, due to accretion, and in the anti-glitches likely represent the effect of reconnection between the faster crust and slower core.
It is possible that the glitch we observe at epoch 56685.8 is a standard glitch.
This would require that the star was spinning \textit{down} prior to the 2014 observations.
Given the spin-down reported between 2014 and 2016, it is indeed possible that a similar spin-down was acting on the NS before that time, leading to the glitch.
Another option would be that the process involved here is due to a high magnetic field, like in magnetars. 
This would agree with the secular spin-down being due to dipole radiation.
We do not have enough data to test this hypothesis.

\subsection{An orbital period derivative?}

As reported in \sref{sec:newdet} and \tref{tab:orbit}, if we select only the data from the 2014 observations and fit the orbital parameters, we obtain a slightly different orbital period than measured including the 2016 dataset.
This might in principle mean that the orbit has shrunk and the orbital period has decreased over time, due for example to the strong mass exchange that the super-Eddington luminosity of the source seems to imply. 
However, the 2014 observations contain a strong timing noise that might have influenced the fit of the orbit reported by B14 and adjusted in this Paper.
We find slightly different solutions depending on the model for the red noise, and \tref{tab:orbit} sums up this variability through larger error bars than the ones calculated from the single fits.

However, if the reported tension between the 2014-only orbital solution and the one with the full data set is true, a single new measurement of the orbital phase in future observations (at this point more than three years since the last one) will be able to measure with high significance a negative orbital period derivative.
The values in \tref{tab:orbit} would imply $\Porbdot \sim-10^{-7}$\,d/d implying an orbital evolution time-scale of 30000\,yrs and a mass exchange almost three orders of magnitude above the Eddington limit (see below).

Following \citet{Rappaport+82}, the expected orbital period derivative for mass transfer from a more massive companion star, neglecting a contribution from gravitational waves, can be estimated through
\begin{multline}\label{eq:masstransfer}
    \dfrac{\dot{a}}{a} = \dfrac{2\Porbdot}{3\Porb} = \\
         -\dfrac{\dot{M}_c}{M_c}
        \left[
            2\left(1 - \frac{\beta}{q}\right) 
            - \frac{1 - \beta}{1 + q}
            - \frac{2\alpha(1-\beta)(1+q)}{q}
        \right] 
\end{multline}
where $M_c$ an $\dot{M}_c$ are the mass and mass loss rate from the companion (negative), $M_p$ an $\dot{M}_p$ the same quantities for the pulsar, $q=M_p/M_c$, $\beta=-\dot{M}_p/\dot{M}_c$, and $\alpha$ the specific angular momentum of the lost mass in units of $2\pi a^2/\Porb$.
In the conservative scenario, $\beta=1$ and \eqref{eq:masstransfer} becomes 
\begin{equation}\label{eq:conservative}
    \dfrac{\dot{a}}{a} = \dfrac{2\Porbdot}{3\Porb} =
         -\dfrac{\dot{M}_c}{M_c}        
        \left[
            2\left(1 - \frac{1}{q}\right)
        \right] 
\end{equation}

In the scenario where all mass is lost in an outflow before accreting on the pulsar, but the outflow is launched from close to the pulsar, the specific angular moment is $j_{\rm outflow}\approx2\pi a_p^2/\Porb$, where $a_p$ is the semimajor axis of the pulsar orbit. 
Therefore $\beta=0$ and $$
\alpha\approx {\left(\frac{a_p}{a}\right)}^2 = {\left(\frac{M_c}{M_p + M_c}\right)}^2 = \frac{1}{(1+q)^2}
$$
and the \eqref{eq:masstransfer} reduces to 
\begin{equation}\label{eq:outflow}
    \dfrac{\dot{a}}{a} = \dfrac{2\Porbdot}{3\Porb} =
         -\dfrac{\dot{M}_c}{M_c}        
        \left[
            2 - \frac{2 + q}{q(1+q)}
        \right] 
\end{equation}
which is a relatively small correction to \eqref{eq:conservative} for small values of $q$.

If the mass transfer in the system is highly super Eddington and no significant mass losses happen close to the donor star, both scenarios predict that the orbit shrinks, with orbital period derivatives of the order 
\begin{align}\label{eq:porbdot}
    \Porbdot &\sim -0.015\,{\left(\frac{M_p}{1.4 M_{\odot}}\right)}^{-1}
        {\left(\frac{-\dot{M}_c}{M_{\odot}/yr}\right)} \,\, \mathrm{s/s}\\
        &\sim -3.5\times10^{-8}\,{\left(\frac{M_p}{1.4 M_{\odot}}\right)}^{-1}
        {\left(\frac{-\dot{M}_c}{100 \dot{M}_{\rm Edd}}\right)} \,\, \mathrm{s/s}
\end{align}
This \Porbdot corresponds to a change of orbital period of $\sim1$\,s/yr, and produces a shift of the orbital phase by a few 100\,s over a few years. 
Again, since $\dot{M}_c$ is negative, also this \Porbdot is negative.
This is easy to observe with standard techniques of pulsar timing. 
Future observations will test this hypothesis by tracking the orbital phase over time.

Finally, the remaining extreme scenario where most mass is lost from the system in form of winds from the donor star, we have
\begin{equation}\label{eq:massloss}
    \dfrac{\dot{a}}{a} = \dfrac{2\Porbdot}{3\Porb} =
         -\dfrac{\dot{M}_c}{M_c}        
            \frac{1}{(1+q)}
\end{equation}
It is easy to show that this scenario would bring the orbit to \textit{expand} instead of shrinking (positive \Porbdot).

It is likely that in a real-life scenario both phenomena should happen, stabilizing the mass transfer over long time scales. 
\citet{fragos_formation_2015} discuss these arguments in detail.
The detection of an orbital period derivative would set strong constraints on the viable models for the evolution of the binary system in \Mtwo.
 
\section{Conclusions}\label{sec:conclusions}
This work characterizes the timing behavior of \Mtwo over two years, showing a number of phenomena that can be used to understand the nature of this remarkable source.
None of these new findings provide definitive information on the nature of the source, but they represent important clues.
Thanks to this work, we now know that:
\begin{itemize}
    \item The pulsar alternates phases with a large spin-up to phases with a large spin-down. Moreover, the spin-up is inconsistent with small values of the fastness parameter. This is a strong indication that the pulsar is close to spin equilibrium.
    \item The pulsations are transient, changing their significance over time, with a tentative positive correlation with the flux of the source.
    This points to an intrinsic mechanism for this pulsed fraction variability rather than, e.g., a change of the background flux from the nearby \Mone.
    \item The neutron star has very strong glitches, probably due to the rapid spin evolution. The observed positive glitch suggests a strong spin-down to have occurred prior to the first detection in 2014.
\end{itemize}
In addition, we now have a very precise orbital solution, that can be used to look for orbital phase derivatives in future observations.
This will be crucial to test the total mass exchange in the system. 

\acknowledgments
MB thanks the Fulbright Visiting Scholar Program for supporting a nine-month visit at Caltech.
MM appreciates support from an STFC Ernest Rutherford Fellowship.
Part of the pulsar search and optimization software was developed in the framework of  CICLOPS -- Citizen Computing Pulsar Search, a project supported by \textit{POR FESR Sardegna 2014 – 2020 Asse 1 Azione 1.1.3} (code RICERCA\_1C-181), call for proposal "Aiuti per Progetti di Ricerca e Sviluppo 2017" managed  by Sardegna Ricerche.
The authors wish to thank Jim Fuller for insightful conversations on stellar pulsations, Georgios Vasilopoulos for comments on the manuscript, Sergei Popov for pointing out alternative models for magnetar evolution, Wynn Ho for pointing out a miscitation about glitch models in the first version posted on the arXiv. Thanks also to Deepto Chakrabarty, Fred Lamb, Juri Poutanen, Alexander Mushtukov and Andrew King for multiple insights on accretion theories and competing models for PULXs.

\vspace{5mm}
\facilities{\nustar \citep{nustar13}, 
            \chandra \citep{chandra02},
            ATNF pulsar catalogue \citep{manchesterAustraliaTelescopeNational2005a}}


\software{HEAsoft \citep{centerheasarcHEAsoftUnifiedRelease2014},
          FTOOLS \citep{blackburnFTOOLSGeneralPackage1999,
                        blackburnFTOOLSFITSData1995a},
          Stingray \citep{huppenkothen_stingray:_2016,
                          huppenkothen-stingray_2019},
          HENDRICS \citep{bachetti_hendrics:_2018},
          astropy \citep{price-whelan_astropy_2018},  
          PINT \citep{luo_pint:_2019},
          PRESTO \citep{ransom_presto:_2011}
          }

\bibliographystyle{yahapj}
\bibliography{papers3,main}

\begin{thebibliography}{}
\providecommand\natexlab[1]{#1}
\providecommand\JournalTitle[1]{#1}

\bibitem[{Andersen \& Ransom(2018)}]{andersen_fourier_2018}
Andersen, B.~C., \& Ransom, S.~M. 2018,
  \href{http://dx.doi.org/10.3847/2041-8213/aad59f}{\JournalTitle{The
  Astrophysical Journal}, 863, L13}

\bibitem[{Archibald {et~al.}(2016)Archibald, Kaspi, Tendulkar, \&
  Scholz}]{archibald_magnetar-like_2016}
Archibald, R.~F., Kaspi, V.~M., Tendulkar, S.~P., \& Scholz, P. 2016,
  \href{http://dx.doi.org/10.3847/2041-8205/829/1/L21}{\JournalTitle{The
  Astrophysical Journal}, 829, L21}

\bibitem[{Bachetti(2018)}]{bachetti_hendrics:_2018}
Bachetti, M. 2018,
  \href{http://adsabs.harvard.edu/abs/2018ascl.soft05019B}{\JournalTitle{Astrophysics
  Source Code Library}, ascl:1805.019}

\bibitem[{Bachetti {et~al.}(2014)Bachetti, Harrison, Walton, Grefenstette,
  Chakrabarty, F{\"u}rst, Barret, Beloborodov, Boggs, Christensen, Craig,
  Fabian, Hailey, Hornschemeier, Kaspi, Kulkarni, Maccarone, Miller, Rana,
  Stern, Tendulkar, Tomsick, Webb, \& Zhang}]{Bachetti+14}
Bachetti, M., Harrison, F.~A., Walton, D.~J., {et~al.} 2014,
  \JournalTitle{Nat.}, 514, 202

\bibitem[{Basko \& Sunyaev(1976)}]{BaskoSunyaev76}
Basko, M.~M., \& Sunyaev, R.~A. 1976, \JournalTitle{MNRAS}, 175, 395

\bibitem[{Begelman {et~al.}(2006)Begelman, King, \&
  Pringle}]{begelman_nature_2006}
Begelman, M.~C., King, A.~R., \& Pringle, J.~E. 2006,
  \href{http://dx.doi.org/10.1111/j.1365-2966.2006.10469.x}{\JournalTitle{Monthly
  Notices of the Royal Astronomical Society}, 370, 399}

\bibitem[{Bildsten \& Brown(1997)}]{Bildsten:1997tp}
Bildsten, L., \& Brown, E.~F. 1997, \JournalTitle{ApJ}, 477, 897

\bibitem[{Blaauw(1961)}]{blaauw_origin_1961}
Blaauw, A. 1961, \JournalTitle{\bain}, 15, 265

\bibitem[{Blackburn(1995)}]{blackburnFTOOLSFITSData1995a}
Blackburn, J.~K. 1995, in Astronomical {{Data Analysis Software}} and {{Systems
  IV}}, Vol.~77, 367

\bibitem[{Blackburn {et~al.}(1999)Blackburn, Shaw, Payne, Hayes, \&
  HEASARC}]{blackburnFTOOLSGeneralPackage1999}
Blackburn, J.~K., Shaw, R.~A., Payne, H.~E., Hayes, J. J.~E., \& HEASARC. 1999,
  \JournalTitle{Astrophysics Source Code Library}, ascl:9912.002

\bibitem[{Boersma(1961)}]{boersma_mathematical_1961}
Boersma, J. 1961, \JournalTitle{bain}, 15, 291

\bibitem[{Brightman {et~al.}(2016)Brightman, Harrison, Walton, {F\"urst},
  Hornschemeier, Zezas, Bachetti, Grefenstette, Ptak, Tendulkar, \&
  Yukita}]{brightman_spectral_2016}
Brightman, M., Harrison, F., Walton, D.~J., {et~al.} 2016,
  \href{http://dx.doi.org/10.3847/0004-637X/816/2/60}{\JournalTitle{ApJ}, 816,
  60}

\bibitem[{Brightman {et~al.}(2019)Brightman, Harrison, Bachetti, Xu, F{\"}rst,
  Walton, Ptak, Yukita, \& Zezas}]{brightman_60_2019}
Brightman, M., Harrison, F.~A., Bachetti, M., {et~al.} 2019,
  \href{http://dx.doi.org/10.3847/1538-4357/ab0215}{\JournalTitle{The
  Astrophysical Journal}, 873, 115}

\bibitem[{Brumback {et~al.}(2018)Brumback, Hickox, Bachetti, Ballhausen,
  {F\"urst}, Pike, Pottschmidt, Tomsick, \& Wilms}]{brumback_discovery_2018}
Brumback, M.~C., Hickox, R.~C., Bachetti, M., {et~al.} 2018,
  \href{http://dx.doi.org/10.3847/2041-8213/aacd13}{\JournalTitle{The
  Astrophysical Journal}, 861, L7}

\bibitem[{Buccheri {et~al.}(1983)Buccheri, Bennett, Bignami, Bloemen,
  Boriakoff, Caraveo, Hermsen, Kanbach, Manchester, Masnou, Mayer-Hasselwander,
  Ozel, Paul, Sacco, Scarsi, \& Strong}]{buccheri_search_1983}
Buccheri, R., Bennett, K., Bignami, G.~F., {et~al.} 1983,
  \href{http://adsabs.harvard.edu/cgi-bin/nph-data_query?bibcode=1983A%26A...128..245B&link_type=ABSTRACT}{\JournalTitle{A\&A},
  128, 245}

\bibitem[{Burderi {et~al.}(2006)Burderi, {di Salvo}, Menna, Riggio, \&
  Papitto}]{burderiOrderChaosSpinup2006}
Burderi, L., {di Salvo}, T., Menna, M.~T., Riggio, A., \& Papitto, A. 2006,
  \href{http://dx.doi.org/10.1086/510666}{\JournalTitle{ApJ}, 653, L133}

\bibitem[{Canal \& Schatzman(1976)}]{canal_non_1976}
Canal, R., \& Schatzman, E. 1976,
  \href{https://ui.adsabs.harvard.edu/abs/1976A%26A....46..229C/abstract}{\JournalTitle{Astronomy
  and Astrophysics}, 46, 229}

\bibitem[{Carpano {et~al.}(2018)Carpano, Haberl, Maitra, \&
  Vasilopoulos}]{carpano_discovery_2018}
Carpano, S., Haberl, F., Maitra, C., \& Vasilopoulos, G. 2018,
  \href{http://dx.doi.org/10.1093/mnrasl/sly030}{\JournalTitle{MNRAS Let.},
  476, L45}

\bibitem[{Casella {et~al.}(2008)Casella, Altamirano, Patruno, Wijnands, \&
  van~der Klis}]{Casella+08}
Casella, P., Altamirano, D., Patruno, A., Wijnands, R. A.~D., \& van~der Klis,
  M. 2008, \JournalTitle{ApJ}, 674, L41

\bibitem[{Conroy {et~al.}(2018)Conroy, Strader, van Dokkum, Dolphin, Weisz,
  Murphy, Dotter, Johnson, \& Cargile}]{conroy_complete_2018}
Conroy, C., Strader, J., van Dokkum, P., {et~al.} 2018,
  \href{http://dx.doi.org/10.3847/1538-4357/aad460}{\JournalTitle{The
  Astrophysical Journal}, 864, 111}

\bibitem[{Dall'Osso {et~al.}(2015)Dall'Osso, Perna, \& Stella}]{Dallosso+15}
Dall'Osso, S., Perna, R., \& Stella, L. 2015, \JournalTitle{MNRAS}, 449, 2144

\bibitem[{Dauser {et~al.}(2017)Dauser, Middleton, \&
  Wilms}]{dauser_modelling_2017}
Dauser, T., Middleton, M., \& Wilms, J. 2017,
  \href{http://dx.doi.org/10.1093/mnras/stw3304}{\JournalTitle{Monthly Notices
  of the Royal Astronomical Society}, 466, 2236}

\bibitem[{Dib \& Kaspi(2014)}]{dib_16_2014}
Dib, R., \& Kaspi, V.~M. 2014,
  \href{http://dx.doi.org/10.1088/0004-637X/784/1/37}{\JournalTitle{{ApJ}},
  784, 37}

\bibitem[{Dib {et~al.}(2009)Dib, Kaspi, \& Gavriil}]{dib_rossi_2009}
Dib, R., Kaspi, V.~M., \& Gavriil, F.~P. 2009,
  \href{http://dx.doi.org/10.1088/0004-637X/702/1/614}{\JournalTitle{The
  Astrophysical Journal}, 702, 614}

\bibitem[{Ek{\c s}i {et~al.}(2015)Ek{\c s}i, Anda{\c c}, {\c
  C}{\i}k{\i}nto{\u{g}}lu, Gen{\c c}ali, G{\"u}ng{\"o}r, \&
  {\"O}ztekin}]{Eksi+15}
Ek{\c s}i, K.~Y., Anda{\c c}, {\.{I}}.~C., {\c C}{\i}k{\i}nto{\u{g}}lu, S.,
  {et~al.} 2015, \JournalTitle{MNRAS Let.}, 448, L40

\bibitem[{Ergma(1993)}]{ergma_accretion_1993}
Ergma, E. 1993, \JournalTitle{\aap}, 273, L38

\bibitem[{Fabrika {et~al.}(2015)Fabrika, Ueda, Vinokurov, Sholukhova, \&
  Shidatsu}]{Fabrika+15}
Fabrika, S., Ueda, Y., Vinokurov, A., Sholukhova, O., \& Shidatsu, M. 2015,
  \JournalTitle{Nature Physics}, 11, 551

\bibitem[{Fragos {et~al.}(2015)Fragos, Linden, Kalogera, \&
  Sklias}]{fragos_formation_2015}
Fragos, T., Linden, T., Kalogera, V., \& Sklias, P. 2015,
  \href{http://dx.doi.org/10.1088/2041-8205/802/1/L5}{\JournalTitle{The
  Astrophysical Journal}, 802, L5}

\bibitem[{Fragos \& McClintock(2015)}]{fragos_origin_2015}
Fragos, T., \& McClintock, J.~E. 2015,
  \href{http://dx.doi.org/10.1088/0004-637X/800/1/17}{\JournalTitle{The
  Astrophysical Journal}, 800, 17}

\bibitem[{Frank {et~al.}(2002)Frank, King, \& Raine}]{FrankKingRaine}
Frank, J., King, A., \& Raine, D.~J. 2002, {Accretion Power in Astrophysics:
  Third Edition} (AA(Louisiana State University), AB(University of Leicester),
  AC(University of Leicester): Accretion Power in Astrophysics)

\bibitem[{Fryer {et~al.}(1999)Fryer, Benz, Herant, \&
  Colgate}]{fryer_what_1999}
Fryer, C., Benz, W., Herant, M., \& Colgate, S.~A. 1999,
  \href{http://dx.doi.org/10.1086/307119}{\JournalTitle{\apj}, 516, 892}

\bibitem[{Fuller(2017)}]{fuller_heartbeat_2017}
Fuller, J. 2017,
  \href{http://dx.doi.org/10.1093/mnras/stx2135}{\JournalTitle{Monthly Notices
  of the Royal Astronomical Society}, 472, 1538}

\bibitem[{F\"urst {et~al.}(2016)F\"urst, Walton, Harrison, Stern, Barret,
  Brightman, Fabian, Grefenstette, Madsen, Middleton, Miller, Pottschmidt,
  Ptak, Rana, \& Webb}]{furst_discovery_2016}
F\"urst, F., Walton, D.~J., Harrison, F.~A., {et~al.} 2016,
  \href{http://dx.doi.org/10.3847/2041-8205/831/2/L14}{\JournalTitle{{ApJL}},
  831, L14}

\bibitem[{Galloway {et~al.}(2006)Galloway, Morgan, Krauss, Kaaret, \&
  Chakrabarty}]{galloway_intermittent_2006}
Galloway, D.~K., Morgan, E.~H., Krauss, M.~I., Kaaret, P., \& Chakrabarty, D.
  2006, \href{http://dx.doi.org/10.1086/510741}{\JournalTitle{ApJ}, 654, L73}

\bibitem[{Galloway {et~al.}(2004)Galloway, Morgan, \&
  Levine}]{galloway_frequency_2004}
Galloway, D.~K., Morgan, E.~H., \& Levine, A.~M. 2004,
  \href{http://dx.doi.org/10.1086/423265}{\JournalTitle{The Astrophysical
  Journal}, 613, 1164}

\bibitem[{Ghosh \& Lamb(1979{\natexlab{a}})}]{GL79a}
Ghosh, P., \& Lamb, F.~K. 1979{\natexlab{a}}, \JournalTitle{ApJ}, 232, 259

\bibitem[{Ghosh \& Lamb(1979{\natexlab{b}})}]{GL79b}
---. 1979{\natexlab{b}}, \JournalTitle{ApJ}, 234, 296

\bibitem[{Gris{\'e} {et~al.}(2011)Gris{\'e}, Kaaret, Pakull, \&
  Motch}]{griseOpticalPropertiesUltraluminous2011}
Gris{\'e}, F., Kaaret, P., Pakull, M.~W., \& Motch, C. 2011,
  \href{http://dx.doi.org/10.1088/0004-637X/734/1/23}{\JournalTitle{ApJ}, 734,
  23}

\bibitem[{Harrison {et~al.}(2013)Harrison, Craig, Christensen, Hailey, Zhang,
  Boggs, Stern, Cook, Forster, Giommi, Grefenstette, Kim, Kitaguchi, Koglin,
  Madsen, Mao, Miyasaka, Mori, Perri, Pivovaroff, Puccetti, Rana, Westergaard,
  Willis, Zoglauer, An, Bachetti, Barri{\`e}re, Bellm, Bhalerao, Brejnholt,
  Fuerst, Liebe, Markwardt, Nynka, Vogel, Walton, Wik, Alexander, Cominsky,
  Hornschemeier, Hornstrup, Kaspi, Madejski, Matt, Molendi, Smith, Tomsick,
  Ajello, Ballantyne, Balokovi{\'c}, Barret, Bauer, Blandford, Brandt,
  Brenneman, Chiang, Chakrabarty, Chenevez, Comastri, Dufour, Elvis, Fabian,
  Farrah, Fryer, Gotthelf, Grindlay, Helfand, Krivonos, Meier, Miller,
  Natalucci, Ogle, Ofek, Ptak, Reynolds, Rigby, Tagliaferri, Thorsett,
  Treister, \& Urry}]{nustar13}
Harrison, F.~A., Craig, W.~W., Christensen, F.~E., {et~al.} 2013,
  \JournalTitle{ApJ}, 770, 103

\bibitem[{HEASARC(2014)}]{centerheasarcHEAsoftUnifiedRelease2014}
HEASARC. 2014, \JournalTitle{Astrophysics Source Code Library}, ascl:1408.004

\bibitem[{Heida {et~al.}(2019)Heida, Harrison, Brightman, F\"rst, Stern, \&
  Walton}]{heida_searching_2019}
Heida, M., Harrison, F.~A., Brightman, M., {et~al.} 2019,
  \href{http://dx.doi.org/10.3847/1538-4357/aafa77}{\JournalTitle{The
  Astrophysical Journal}, 871, 231}

\bibitem[{Huppenkothen {et~al.}(2016)Huppenkothen, Bachetti, Stevens, Migliari,
  \& Balm}]{huppenkothen_stingray:_2016}
Huppenkothen, D., Bachetti, M., Stevens, A.~L., Migliari, S., \& Balm, P. 2016,
  \href{http://adsabs.harvard.edu/cgi-bin/nph-data_query?bibcode=2016ascl.soft08001H&link_type=EJOURNAL}{\JournalTitle{Astrophysics
  Source Code Library}, ascl:1608.001}

\bibitem[{{Huppenkothen} {et~al.}(2019){Huppenkothen}, {Bachetti}, {Stevens},
  {Migliari}, {Balm}, {Hammad}, {Khan}, {Mishra}, {Rashid}, {Sharma}, {Blanco},
  \& {Ribeiro}}]{huppenkothen-stingray_2019}
{Huppenkothen}, D., {Bachetti}, M., {Stevens}, A.~L., {et~al.} 2019,
  \JournalTitle{arXiv e-prints}, arXiv:1901.07681

\bibitem[{Hut \& Paczynski(1984)}]{hut_effects_1984}
Hut, P., \& Paczynski, B. 1984,
  \href{http://dx.doi.org/10.1086/162450}{\JournalTitle{The Astrophysical
  Journal}, 284, 675}

\bibitem[{Igoshev \& Popov(2018)}]{igoshevHowMakeMature2018}
Igoshev, A.~P., \& Popov, S.~B. 2018,
  \href{http://dx.doi.org/10.1093/mnras/stx2573}{\JournalTitle{Monthly Notices
  of the Royal Astronomical Society}, 473, 3204}

\bibitem[{Israel {et~al.}(2017{\natexlab{a}})Israel, Belfiore, Stella,
  Esposito, Casella, De~Luca, Marelli, Papitto, Perri, Puccetti, Castillo,
  Salvetti, Tiengo, Zampieri, D’Agostino, Greiner, Haberl, Novara,
  Salvaterra, Turolla, Watson, Wilms, \& Wolter}]{israel_accreting_2017}
Israel, G.~L., Belfiore, A., Stella, L., {et~al.} 2017{\natexlab{a}},
  \href{http://dx.doi.org/10.1126/science.aai8635}{\JournalTitle{Science}, 355,
  817}

\bibitem[{Israel {et~al.}(2017{\natexlab{b}})Israel, Papitto, Esposito, Stella,
  Zampieri, Belfiore, Rodríguez~Castillo, de~Luca, Tiengo, Haberl, Greiner,
  Salvaterra, Sandrelli, \& Lisini}]{israel_discovery_2017}
Israel, G.~L., Papitto, A., Esposito, P., {et~al.} 2017{\natexlab{b}},
  \href{http://dx.doi.org/10.1093/mnrasl/slw218}{\JournalTitle{{MNRAS} Let.},
  466, L48}

\bibitem[{Jiang {et~al.}(2018)Jiang, Cantiello, Bildsten, Quataert, Blaes, \&
  Stone}]{jiang_three_2018}
Jiang, Y.-F., Cantiello, M., Bildsten, L., {et~al.} 2018,
  \href{https://ui.adsabs.harvard.edu/abs/2018arXiv180910187J/abstract}{\JournalTitle{arXiv
  e-prints}, arXiv:1809.10187}

\bibitem[{Kaaret {et~al.}(2017)Kaaret, Feng, \&
  Roberts}]{kaaretUltraluminousXRaySources2017}
Kaaret, P., Feng, H., \& Roberts, T.~P. 2017,
  \href{http://dx.doi.org/10.1146/annurev-astro-091916-055259}{\JournalTitle{Annual
  Review of Astronomy and Astrophysics}, 55, 303}

\bibitem[{Kalogera(1998)}]{kalogera_formation_1998}
Kalogera, V. 1998, \href{http://dx.doi.org/10.1086/305086}{\JournalTitle{\apj},
  493, 368}

\bibitem[{King {et~al.}(2017)King, Lasota, \&
  Klu{\'z}niak}]{kingPulsingULXsTip2017}
King, A., Lasota, J.-P., \& Klu{\'z}niak, W. 2017,
  \href{http://dx.doi.org/10.1093/mnrasl/slx020}{\JournalTitle{MNRAS Let.},
  468, L59}

\bibitem[{King(2008)}]{kingAccretionRatesBeaming2008}
King, A.~R. 2008,
  \href{http://dx.doi.org/10.1111/j.1745-3933.2008.00444.x}{\JournalTitle{MNRAS
  Let.}, 385, L113}

\bibitem[{King(2009)}]{kingMassesBeamingEddington2009}
---. 2009,
  \href{http://dx.doi.org/10.1111/j.1745-3933.2008.00594.x}{\JournalTitle{MNRAS
  Let.}, 393, L41}

\bibitem[{King {et~al.}(2001)King, Pringle, \& Wickramasinghe}]{king_type_2001}
King, A.~R., Pringle, J.~E., \& Wickramasinghe, D.~T. 2001,
  \href{http://dx.doi.org/10.1046/j.1365-8711.2001.04184.x}{\JournalTitle{\mnras},
  320, L45}

\bibitem[{Knigge {et~al.}(2011)Knigge, Coe, \& Podsiadlowski}]{knigge_two_2011}
Knigge, C., Coe, M.~J., \& Podsiadlowski, P. 2011,
  \href{http://dx.doi.org/10.1038/nature10529}{\JournalTitle{\nat}, 479, 372}

\bibitem[{Koliopanos {et~al.}(2017)Koliopanos, Vasilopoulos, Godet, Bachetti,
  Webb, \& Barret}]{koliopanos_ulx_2017}
Koliopanos, F., Vasilopoulos, G., Godet, O., {et~al.} 2017,
  \href{http://dx.doi.org/10.1051/0004-6361/201730922}{\JournalTitle{A\&A},
  608, A47}

\bibitem[{Kosec {et~al.}(2018)Kosec, Pinto, Walton, Fabian, Bachetti,
  Brightman, Fürst, \& Grefenstette}]{kosec_evidence_2018}
Kosec, P., Pinto, C., Walton, D.~J., {et~al.} 2018,
  \href{http://dx.doi.org/10.1093/mnras/sty1626}{\JournalTitle{Monthly Notices
  of the Royal Astronomical Society}, 479, 3978}

\bibitem[{Kozai(1962)}]{kozai_secular_1962}
Kozai, Y. 1962, \href{http://dx.doi.org/10.1086/108790}{\JournalTitle{The
  Astronomical Journal}, 67, 591}

\bibitem[{Kretschmar {et~al.}(2004)Kretschmar, Wilms, Staubert, Kreykenbohm, \&
  Heindl}]{kretschmar_xmm-newton_2004}
Kretschmar, P., Wilms, J., Staubert, R., Kreykenbohm, I., \& Heindl, W.~A.
  2004, \href{http://adsabs.harvard.edu/abs/2004ESASP.552..329K}{in Proceedings
  of the 5th {INTEGRAL} {Workshop} on the {INTEGRAL} {Universe} ({ESA}
  {SP}-552). 16-20 {February} 2004}, 329

\bibitem[{Lewin {et~al.}(1988)Lewin, van Paradijs, \& van~der Klis}]{Lewin+88}
Lewin, W. H.~G., van Paradijs, J., \& van~der Klis, M. 1988,
  \JournalTitle{SSRv}, 46, 273

\bibitem[{Lidov(1962)}]{lidov_evolution_1962}
Lidov, M.~L. 1962,
  \href{http://dx.doi.org/10.1016/0032-0633(62)90129-0}{\JournalTitle{Planetary
  and Space Science}, 9, 719}

\bibitem[{Luo {et~al.}(2019)Luo, Ransom, Demorest, van Haasteren, Ray, Stovall,
  Bachetti, Archibald, Kerr, Colen, \& Jenet}]{luo_pint:_2019}
Luo, J., Ransom, S., Demorest, P., {et~al.} 2019,
  \href{http://adsabs.harvard.edu/abs/2019ascl.soft02007L}{\JournalTitle{Astrophysics
  Source Code Library}, ascl:1902.007}

\bibitem[{Maccarone {et~al.}(2010)Maccarone, Kundu, Zepf, \&
  Rhode}]{maccarone_fading_2010}
Maccarone, T.~J., Kundu, A., Zepf, S.~E., \& Rhode, K.~L. 2010,
  \href{http://dx.doi.org/10.1111/j.1745-3933.2010.00953.x}{\JournalTitle{Monthly
  Notices of the Royal Astronomical Society}, 409, L84}

\bibitem[{Manchester(2018)}]{manchester_pulsar_2018}
Manchester, R.~N. 2018,
  \href{http://arxiv.org/abs/1801.04332}{\JournalTitle{{arXiv}:1801.04332
  [astro-ph]}}, \href{http://arxiv.org/abs/1801.04332}{{\sffamily 1801.04332}}

\bibitem[{Manchester {et~al.}(2005)Manchester, Hobbs, Teoh, \&
  Hobbs}]{manchesterAustraliaTelescopeNational2005a}
Manchester, R.~N., Hobbs, G.~B., Teoh, A., \& Hobbs, M. 2005,
  \href{http://dx.doi.org/10.1086/428488}{\JournalTitle{The Astronomical
  Journal}, 129, 1993}

\bibitem[{Mellah {et~al.}(2019)Mellah, Sundqvist, \&
  Keppens}]{mellahWindRocheLobe2019}
Mellah, I.~E., Sundqvist, J.~O., \& Keppens, R. 2019,
  \href{http://dx.doi.org/10.1051/0004-6361/201834543}{\JournalTitle{A\&A},
  622, L3}

\bibitem[{Middleton {et~al.}(2019{\natexlab{a}})Middleton, Brightman, Pintore,
  Bachetti, Fabian, F{\"u}rst, \& Walton}]{middletonMagneticFieldM512019}
Middleton, M.~J., Brightman, M., Pintore, F., {et~al.} 2019{\natexlab{a}},
  \href{http://dx.doi.org/10.1093/mnras/stz436}{\JournalTitle{Monthly Notices
  of the Royal Astronomical Society}, 486, 2}

\bibitem[{Middleton {et~al.}(2019{\natexlab{b}})Middleton, Fragile, Ingram, \&
  Roberts}]{middletonLenseThirringTimingaccretionPlane2019}
Middleton, M.~J., Fragile, P.~C., Ingram, A., \& Roberts, T.~P.
  2019{\natexlab{b}},
  \href{http://dx.doi.org/10.1093/mnras/stz2005}{\JournalTitle{Monthly Notices
  of the Royal Astronomical Society}, 1946}

\bibitem[{Middleton {et~al.}(2015{\natexlab{a}})Middleton, Heil, Pintore,
  Walton, \& Roberts}]{middleton_spectral-timing_2015}
Middleton, M.~J., Heil, L., Pintore, F., Walton, D.~J., \& Roberts, T.~P.
  2015{\natexlab{a}},
  \href{http://dx.doi.org/10.1093/mnras/stu2644}{\JournalTitle{MNRAS}, 447,
  3243}

\bibitem[{Middleton \& King(2017)}]{middleton_predicting_2017}
Middleton, M.~J., \& King, A. 2017,
  \href{http://dx.doi.org/10.1093/mnrasl/slx079}{\JournalTitle{Monthly Notices
  of the Royal Astronomical Society}, 470, L69}

\bibitem[{Middleton {et~al.}(2015{\natexlab{b}})Middleton, Walton, Fabian,
  Roberts, Heil, Pinto, Anderson, \&
  Sutton}]{middletonDiagnosingAccretionFlow2015}
Middleton, M.~J., Walton, D.~J., Fabian, A., {et~al.} 2015{\natexlab{b}},
  \href{http://dx.doi.org/10.1093/mnras/stv2214}{\JournalTitle{MNRAS}, 454,
  3134}

\bibitem[{Mohamed \& Podsiadlowski(2007)}]{mohamedWindRocheLobeOverflow2007}
Mohamed, S., \& Podsiadlowski, P. 2007, \JournalTitle{15th European Workshop on
  White Dwarfs}, 372, 397

\bibitem[{Molkov {et~al.}(2017)Molkov, Lutovinov, Falanga, Tsygankov, \&
  Bozzo}]{molkov_near-periodical_2017}
Molkov, S., Lutovinov, A., Falanga, M., Tsygankov, S., \& Bozzo, E. 2017,
  \href{http://dx.doi.org/10.1093/mnras/stw2429}{\JournalTitle{Monthly Notices
  of the Royal Astronomical Society}, 464, 2039}

\bibitem[{Mushtukov {et~al.}(2019)Mushtukov, Ingram, Middleton, Nagirner, \&
  {van der Klis}}]{mushtukovTimingPropertiesULX2019}
Mushtukov, A.~A., Ingram, A., Middleton, M., Nagirner, D.~I., \& {van der
  Klis}, M. 2019,
  \href{http://dx.doi.org/10.1093/mnras/sty3525}{\JournalTitle{Monthly Notices
  of the Royal Astronomical Society}, 484, 687}

\bibitem[{Mushtukov {et~al.}(2017)Mushtukov, Suleimanov, Tsygankov, \&
  Ingram}]{mushtukovOpticallyThickEnvelopes2017}
Mushtukov, A.~A., Suleimanov, V.~F., Tsygankov, S.~S., \& Ingram, A. 2017,
  \href{http://dx.doi.org/10.1093/mnras/stx141}{\JournalTitle{Monthly Notices
  of the Royal Astronomical Society}, 467, 1202}

\bibitem[{Mushtukov {et~al.}(2015)Mushtukov, Suleimanov, Tsygankov, \&
  Poutanen}]{Mushtukov+15}
Mushtukov, A.~A., Suleimanov, V.~F., Tsygankov, S.~S., \& Poutanen, J. 2015,
  \JournalTitle{arXiv}, 3600

\bibitem[{Nomoto(1987)}]{nomoto_evolution_1987}
Nomoto, K. 1987, \href{http://dx.doi.org/10.1086/165716}{\JournalTitle{\apj},
  322, 206}

\bibitem[{Pasham \& Strohmayer(2013)}]{pashamCan62Day2013}
Pasham, D.~R., \& Strohmayer, T.~E. 2013,
  \href{http://dx.doi.org/10.1088/2041-8205/774/2/L16}{\JournalTitle{ApJ}, 774,
  L16}

\bibitem[{Perna {et~al.}(2006{\natexlab{a}})Perna, Bozzo, \&
  Stella}]{pernaSpinupSpindownTransitions2006}
Perna, R., Bozzo, E., \& Stella, L. 2006{\natexlab{a}},
  \href{http://dx.doi.org/10.1086/499227}{\JournalTitle{The Astrophysical
  Journal}, 639, 363}

\bibitem[{Perna {et~al.}(2006{\natexlab{b}})Perna, Bozzo, \&
  Stella}]{perna_spin-up/spin-down_2006}
---. 2006{\natexlab{b}},
  \href{http://dx.doi.org/10.1086/499227}{\JournalTitle{The Astrophysical
  Journal}, 639, 363}

\bibitem[{Pfahl {et~al.}(2002)Pfahl, Rappaport, Podsiadlowski, \&
  Spruit}]{pfahl_new_2002}
Pfahl, E., Rappaport, S., Podsiadlowski, P., \& Spruit, H. 2002,
  \href{http://dx.doi.org/10.1086/340794}{\JournalTitle{\apj}, 574, 364}

\bibitem[{Pike {et~al.}(2019)Pike, Harrison, Bachetti, Brumback, {F\"urst},
  Madsen, Pottschmidt, Tomsick, \& Wilms}]{pike_observing_2019}
Pike, S.~N., Harrison, F.~A., Bachetti, M., {et~al.} 2019,
  \href{https://ui.adsabs.harvard.edu/#abs/arXiv:1903.06306}{\JournalTitle{arXiv
  e-prints}, arXiv:1903.06306}

\bibitem[{Pinto {et~al.}(2016)Pinto, Fabian, Middleton, \&
  Walton}]{pinto_ultrafast_2016}
Pinto, C., Fabian, A., Middleton, M., \& Walton, D. 2016,
  \href{http://adsabs.harvard.edu/abs/2016arXiv161100623P}{\JournalTitle{arXiv},
  arXiv:1611.00623}, arXiv: 1611.00623

\bibitem[{Pintore {et~al.}(2017)Pintore, Zampieri, Stella, Wolter, Mereghetti,
  \& Israel}]{pintore_pulsator-like_2017}
Pintore, F., Zampieri, L., Stella, L., {et~al.} 2017,
  \href{http://dx.doi.org/10.3847/1538-4357/836/1/113}{\JournalTitle{The
  Astrophysical Journal}, 836, 113}

\bibitem[{Poelarends {et~al.}(2008)Poelarends, Herwig, Langer, \&
  Heger}]{poelarends_supernova_2008}
Poelarends, A. J.~T., Herwig, F., Langer, N., \& Heger, A. 2008,
  \href{http://dx.doi.org/10.1086/520872}{\JournalTitle{\apj}, 675, 614}

\bibitem[{Poutanen {et~al.}(2007)Poutanen, Lipunova, Fabrika, Butkevich, \&
  Abolmasov}]{Poutanen+07}
Poutanen, J., Lipunova, G., Fabrika, S., Butkevich, A.~G., \& Abolmasov, P.
  2007, \JournalTitle{MNRAS}, 377, 1187

\bibitem[{Price-Whelan {et~al.}(2018)Price-Whelan, , Günther, Lim, Crawford,
  Conseil, Shupe, Craig, Dencheva, Ginsburg, VanderPlas, Bradley,
  Pérez-Suárez, de~Val-Borro, Aldcroft, Cruz, Robitaille, Tollerud, Ardelean,
  Babej, Bachetti, Bakanov, Bamford, Barentsen, Barmby, Baumbach, Berry,
  Biscani, Boquien, Bostroem, Bouma, Brammer, Bray, Breytenbach, Buddelmeijer,
  Burke, Calderone, Rodríguez, Cara, Cardoso, Cheedella, Copin, Crichton,
  DÁvella, Deil, Depagne, Dietrich, Donath, Droettboom, Earl, Erben, Fabbro,
  Ferreira, Finethy, Fox, Garrison, Gibbons, Goldstein, Gommers, Greco,
  Greenfield, Groener, Grollier, Hagen, Hirst, Homeier, Horton, Hosseinzadeh,
  Hu, Hunkeler, , Jain, Jenness, Kanarek, Kendrew, Kern, Kerzendorf, Khvalko,
  King, Kirkby, Kulkarni, Kumar, Lee, Lenz, Littlefair, Ma, Macleod,
  Mastropietro, McCully, Montagnac, Morris, Mueller, Mumford, Muna, Murphy,
  Nelson, Nguyen, Ninan, Nöthe, Ogaz, Oh, Parejko, Parley, Pascual, Patil,
  Patil, Plunkett, Prochaska, Rastogi, Janga, Sabater, Sakurikar, Seifert,
  Sherbert, Sherwood-Taylor, Shih, Sick, Silbiger, Singanamalla, {Singer, L.
  P.}, Sladen, Sooley, Sornarajah, Streicher, Teuben, Thomas, Tremblay, Turner,
  Terrón, van Kerkwijk, de~la Vega, Watkins, Weaver, Whitmore, Woillez, \&
  Zabalza}]{price-whelan_astropy_2018}
Price-Whelan, A.~M., , Günther, H.~M., {et~al.} 2018,
  \href{http://arxiv.org/abs/1801.02634v1}{\JournalTitle{arXiv},
  arXiv:1801.02634}, arXiv: 1801.02634

\bibitem[{Ransom(2011)}]{ransom_presto:_2011}
Ransom, S. 2011,
  \href{http://adsabs.harvard.edu/abs/2011ascl.soft07017R}{\JournalTitle{Astrophysics
  Source Code Library}, ascl:1107.017}

\bibitem[{Ransom(2001)}]{ransom_new_2001}
Ransom, S.~M. 2001, PhD thesis, Harvard University

\bibitem[{Ransom {et~al.}(2002)Ransom, Eikenberry, \& Middleditch}]{Ransom+02}
Ransom, S.~M., Eikenberry, S.~S., \& Middleditch, J. 2002, \JournalTitle{The
  Astronomical Journal}, 124, 1788

\bibitem[{Rappaport \& Joss(1977)}]{rappaportAccretionTorquesXray1977}
Rappaport, S., \& Joss, P.~C. 1977,
  \href{http://dx.doi.org/10.1038/266683a0}{\JournalTitle{Nat.}, 266, 683}

\bibitem[{Rappaport {et~al.}(1982)Rappaport, Joss, \& Webbink}]{Rappaport+82}
Rappaport, S., Joss, P.~C., \& Webbink, R.~F. 1982, \JournalTitle{ApJ}, 254,
  616

\bibitem[{Ray {et~al.}(2018)Ray, Guillot, Ho, Kerr, Enoto, Gendreau,
  Arzoumanian, Altamirano, Bogdanov, Campion, Chakrabarty, Jaisawal, Kozon,
  Malacaria, Strohmayer, \& Wolff}]{ray_anti-glitches_2018}
Ray, P.~S., Guillot, S., Ho, W. C.~G., {et~al.} 2018,
  \href{https://ui.adsabs.harvard.edu/#abs/arXiv:1811.09218}{\JournalTitle{{arXiv}
  e-prints}, arXiv:1811.09218}

\bibitem[{Samus' {et~al.}(2017)Samus', Kazarovets, Durlevich, Kireeva, \&
  Pastukhova}]{samus_general_2017}
Samus', N.~N., Kazarovets, E.~V., Durlevich, O.~V., Kireeva, N.~N., \&
  Pastukhova, E.~N. 2017,
  \href{http://dx.doi.org/10.1134/S1063772917010085}{\JournalTitle{Astronomy
  Reports}, 61, 80}

\bibitem[{Sana {et~al.}(2012)Sana, de~Mink, de~Koter, Langer, Evans, Gieles,
  Gosset, Izzard, Le~Bouquin, \& Schneider}]{sana_binary_2012}
Sana, H., de~Mink, S.~E., de~Koter, A., {et~al.} 2012,
  \href{http://dx.doi.org/10.1126/science.1223344}{\JournalTitle{Science}, 337,
  444}

\bibitem[{Sathyaprakash {et~al.}(2019)Sathyaprakash, Roberts, Walton, Fuerst,
  Bachetti, Pinto, Alston, Earnshaw, Fabian, Middleton, \&
  Soria}]{sathyaprakashDiscoveryWeakCoherent2019}
Sathyaprakash, R., Roberts, T.~P., Walton, D.~J., {et~al.} 2019,
  \href{http://dx.doi.org/10.1093/mnrasl/slz086}{\JournalTitle{Monthly Notices
  of the Royal Astronomical Society}, L104}

\bibitem[{Savitzky \& Golay(1964)}]{savitzky_smoothing_1964}
Savitzky, A., \& Golay, M. J.~E. 1964,
  \href{http://dx.doi.org/10.1021/ac60214a047}{\JournalTitle{Anal. Chem.}, 36,
  1627}

\bibitem[{Serim {et~al.}(2017)Serim, {\c S}ahiner, {\c C}erri-Serim, {\.I}nam,
  \& Baykal}]{serimDiscoveryGlitchAccretionpowered2017}
Serim, M.~M., {\c S}ahiner, {\c S}., {\c C}erri-Serim, D., {\.I}nam, S.~{\c
  C}., \& Baykal, A. 2017,
  \href{http://dx.doi.org/10.1093/mnras/stx1771}{\JournalTitle{Mon Not R Astron
  Soc}, 471, 4982}

\bibitem[{Shakura \& Sunyaev(1973)}]{SS73}
Shakura, N.~I., \& Sunyaev, R.~A. 1973, \JournalTitle{A{\&}A}, 24, 337

\bibitem[{Skinner {et~al.}(1982)Skinner, Bedford, Elsner, Leahy, Weisskopf, \&
  Grindlay}]{skinner_discovery_1982}
Skinner, G.~K., Bedford, D.~K., Elsner, R.~F., {et~al.} 1982,
  \href{http://dx.doi.org/10.1038/297568a0}{\JournalTitle{Nature}, 297, 568}

\bibitem[{Taylor(1992)}]{Taylor92}
Taylor, J.~H. 1992, \JournalTitle{Philosophical Transactions: Physical Sciences
  and Engineering}, 341, 117

\bibitem[{Taylor \& Weisberg(1982)}]{Taylor+82}
Taylor, J.~H., \& Weisberg, J.~M. 1982, \JournalTitle{ApJ}, 253, 908

\bibitem[{Tsygankov {et~al.}(2015)Tsygankov, Mushtukov, Suleimanov, \&
  Poutanen}]{Tsygankov+15}
Tsygankov, S.~S., Mushtukov, A.~A., Suleimanov, V.~F., \& Poutanen, J. 2015,
  \JournalTitle{arXiv}, 8288

\bibitem[{Tsygankov {et~al.}(2016)Tsygankov, Mushtukov, Suleimanov, \&
  Poutanen}]{tsygankovPropellerEffectAction2016}
---. 2016, \href{http://dx.doi.org/10.1093/mnras/stw046}{\JournalTitle{MNRAS},
  457, 1101}

\bibitem[{Usov(1992)}]{usov_millisecond_1992}
Usov, V.~V. 1992,
  \href{http://dx.doi.org/10.1038/357472a0}{\JournalTitle{\nat}, 357, 472}

\bibitem[{Vasilopoulos {et~al.}(2018)Vasilopoulos, Haberl, Carpano, \&
  Maitra}]{vasilopoulos_ngc_2018}
Vasilopoulos, G., Haberl, F., Carpano, S., \& Maitra, C. 2018,
  \href{http://dx.doi.org/10.1051/0004-6361/201833442}{\JournalTitle{Astronomy
  and Astrophysics}, 620, L12}

\bibitem[{Vasilopoulos {et~al.}(2019)Vasilopoulos, Petropoulou, Koliopanos,
  Ray, Bailyn, Haberl, \& Gendreau}]{vasilopoulosNGC300ULX12019}
Vasilopoulos, G., Petropoulou, M., Koliopanos, F., {et~al.} 2019,
  \JournalTitle{arXiv e-prints}, arXiv:1905.03740

\bibitem[{Walton {et~al.}(2018)Walton, F\"urst, Heida, Harrison, Barret, Stern,
  Bachetti, Brightman, Fabian, \& Middleton}]{walton_evidence_2018}
Walton, D.~J., F\"urst, F., Heida, M., {et~al.} 2018,
  \href{http://dx.doi.org/10.3847/1538-4357/aab610}{\JournalTitle{The
  Astrophysical Journal}, 856, 128}

\bibitem[{Wang(1995)}]{wang_torque_1995}
Wang, Y.-M. 1995, \href{http://dx.doi.org/10.1086/309649}{\JournalTitle{ApJL},
  449, L153}

\bibitem[{Weisskopf {et~al.}(2002)Weisskopf, Brinkman, Canizares, Garmire,
  Murray, \& {van Speybroeck, L. P.}}]{chandra02}
Weisskopf, M.~C., Brinkman, B., Canizares, C., {et~al.} 2002, \JournalTitle{The
  Publications of the Astronomical Society of the Pacific}, 114, 1

\bibitem[{Wiktorowicz {et~al.}(2018)Wiktorowicz, Lasota, Middleton, \&
  Belczynski}]{wiktorowicz_observed_2018}
Wiktorowicz, G., Lasota, J.-P., Middleton, M., \& Belczynski, K. 2018,
  \href{https://ui.adsabs.harvard.edu/abs/2018arXiv181108998W/abstract}{\JournalTitle{arXiv
  e-prints}, arXiv:1811.08998}

\end{thebibliography}

\appendix
\section{On the detection limits used in the pulsation search.}
\begin{figure*}
\centering
\gridline{\fig{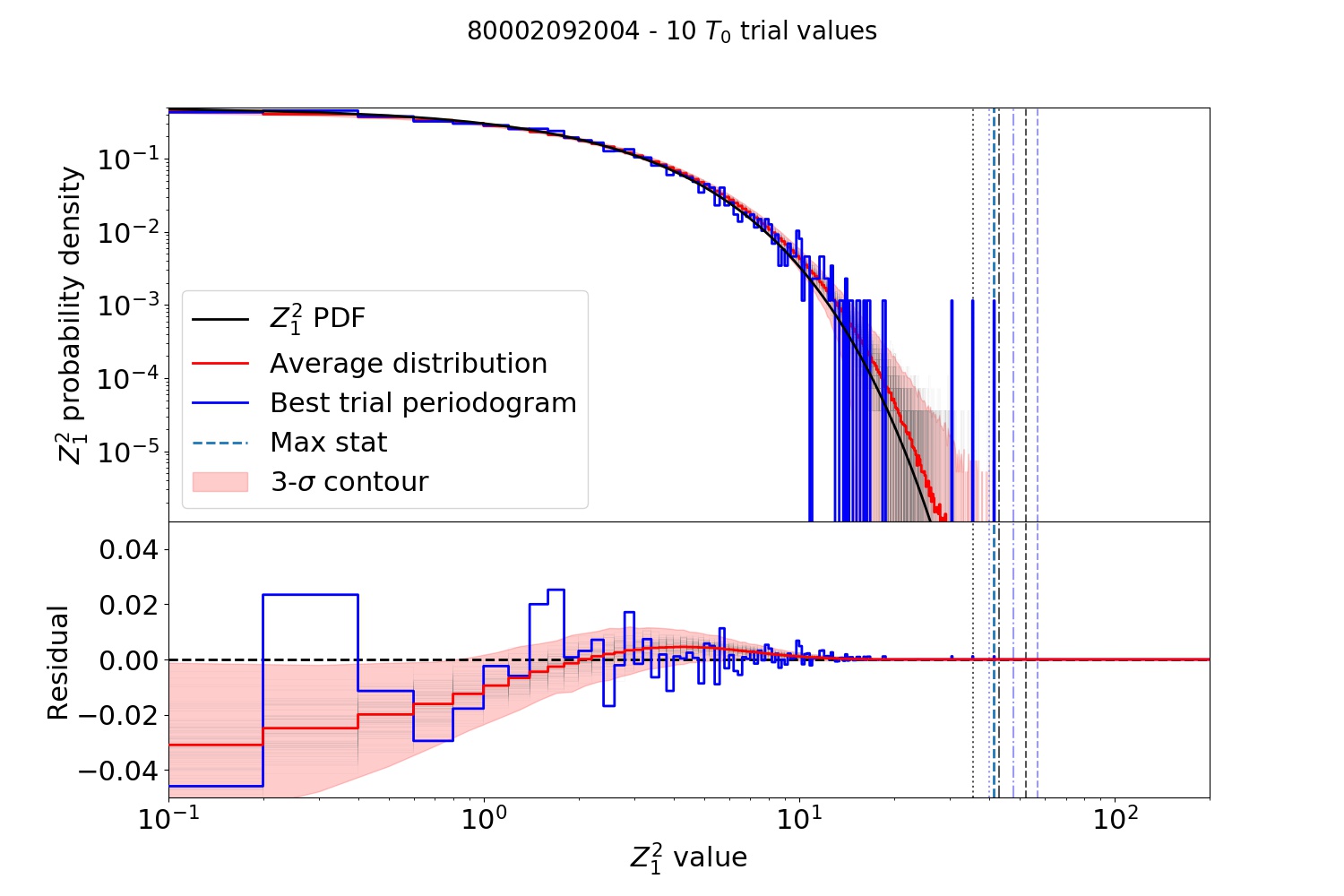}{0.48\textwidth}{(a)}
          \fig{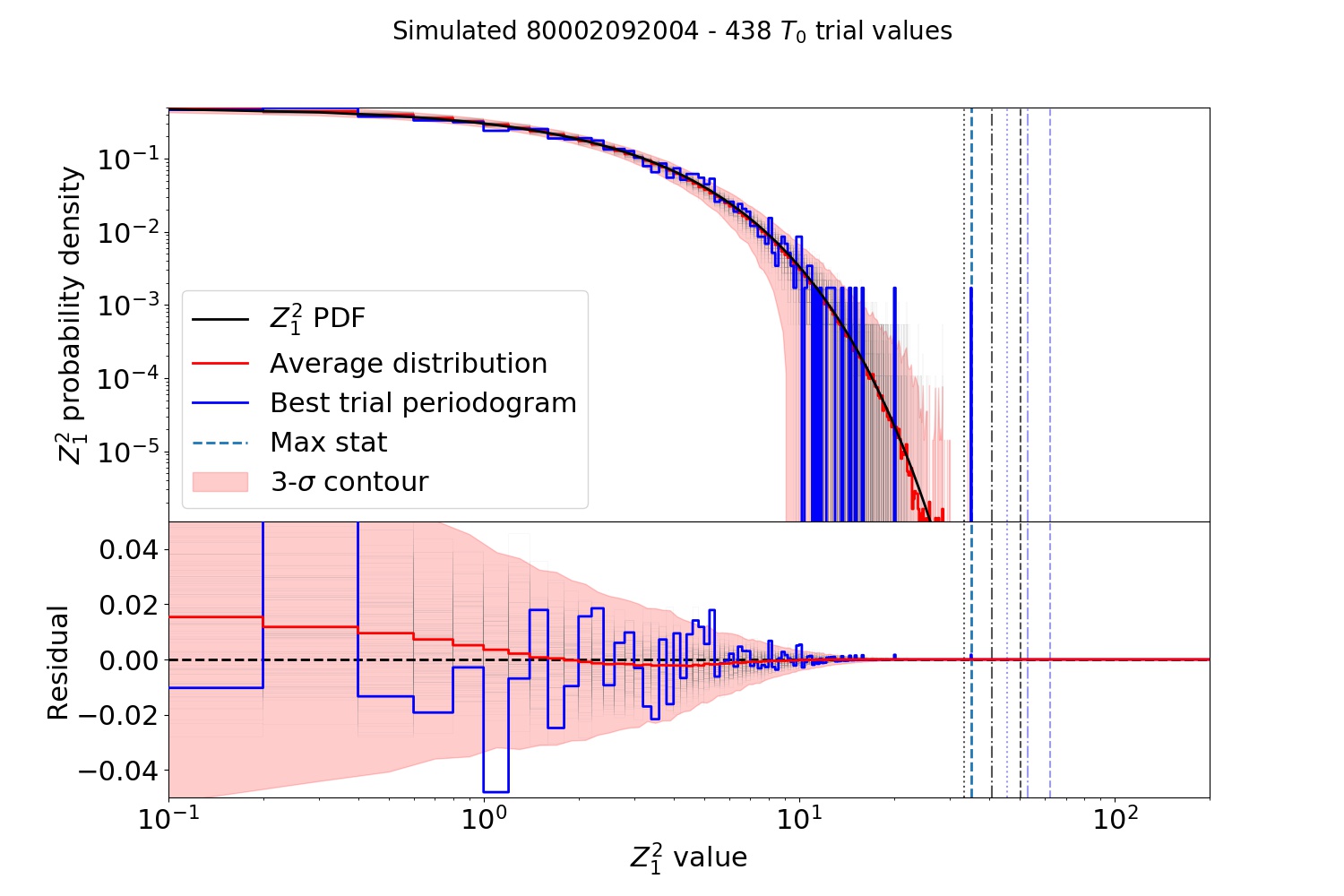}{0.48\textwidth}{(b)}}
\gridline{\fig{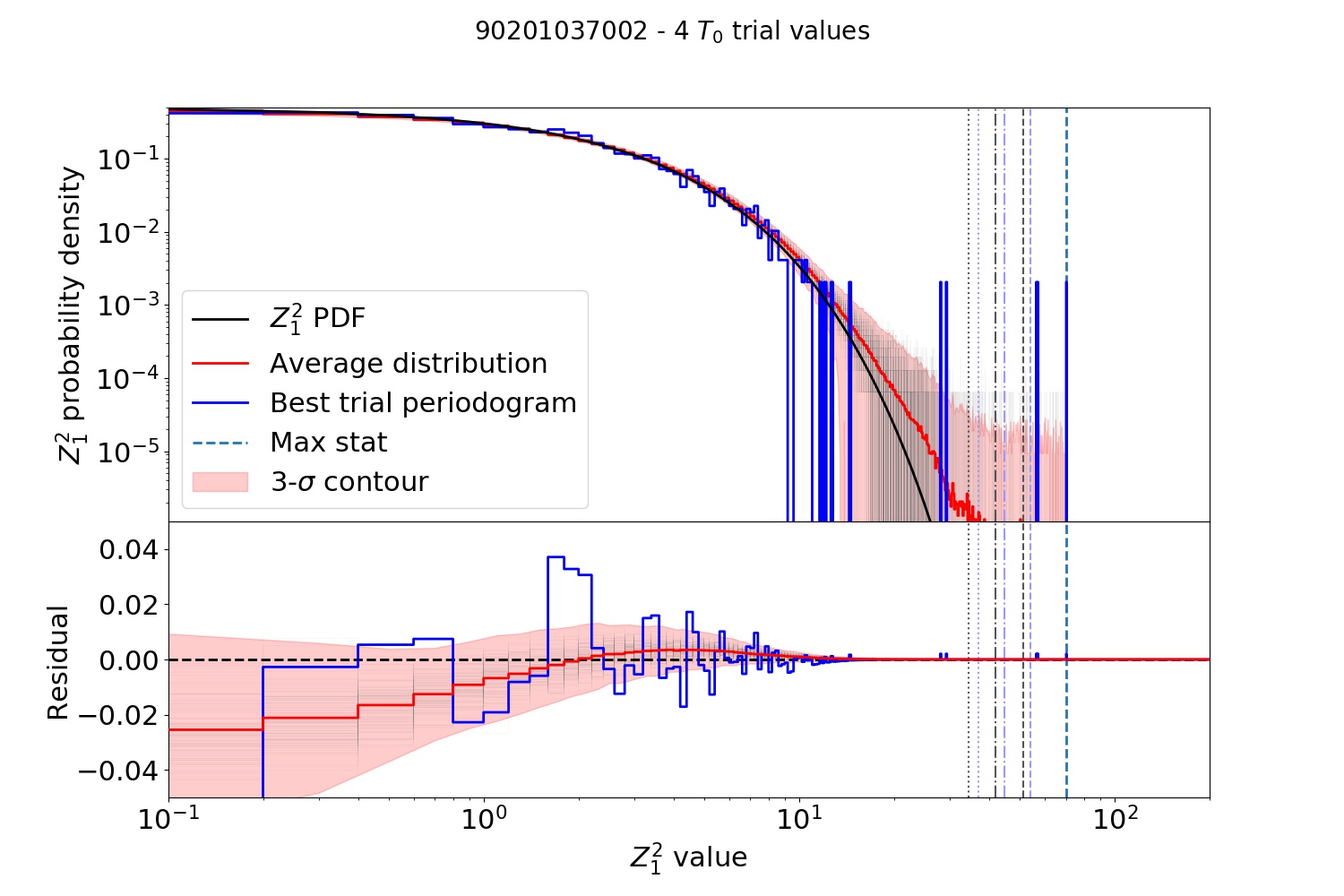}{0.48\textwidth}{(c)}
          \fig{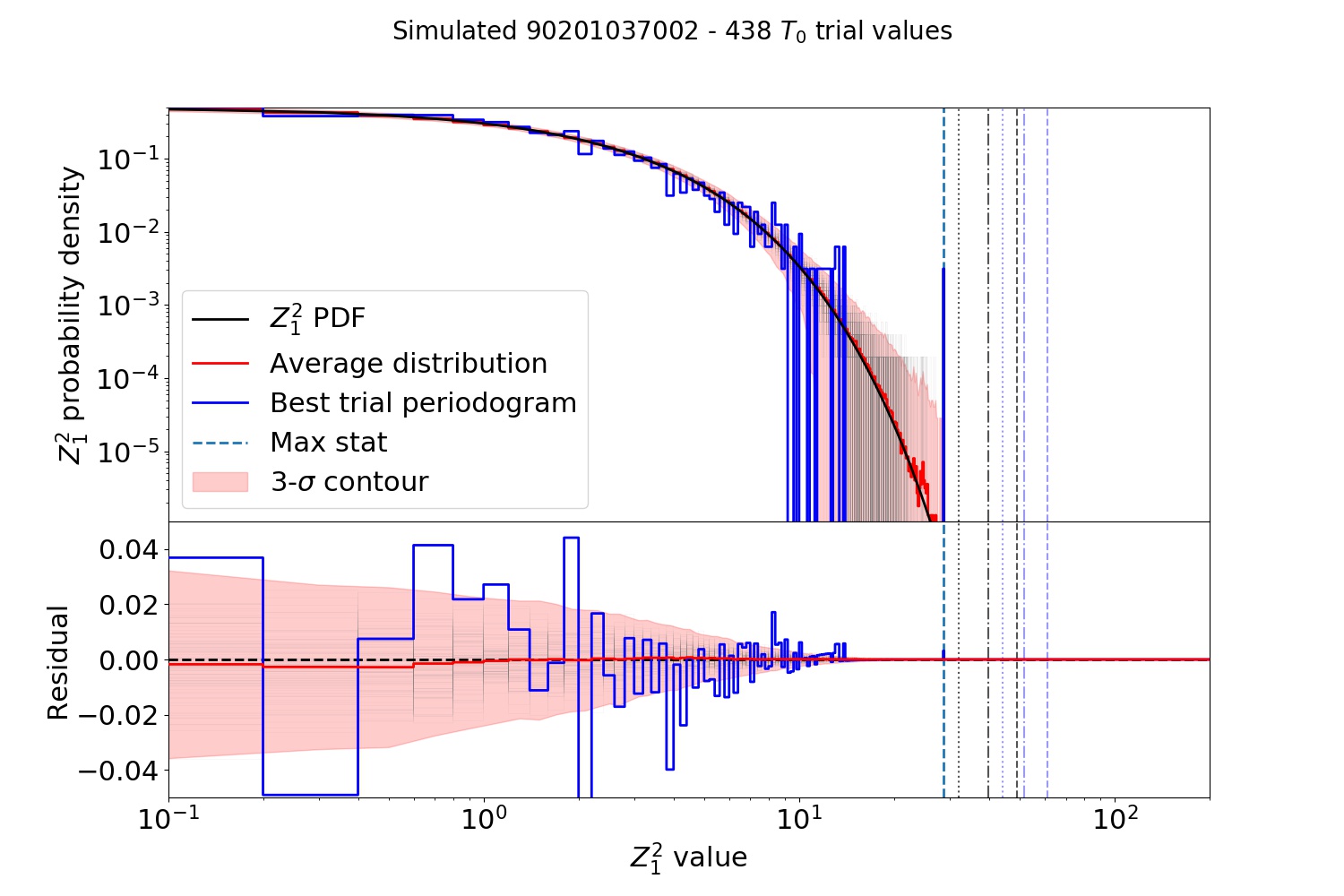}{0.48\textwidth}{(d)}}
\caption{The statistical distribution of $Z_1^2$ values in our blind search of pulsations, using the orbital solution in \tref{tab:orbit} and a search over $T_0$ values and $f,\dot{f}$.
The dotted, dash-dotted and dashed vertical lines show the 3, 4 and 5-$\sigma$ levels from the theoretical distribution, for the case of a single search over $f-\dot{f}$, and accounting for the given number of trial values of $T_0$. 
a/b show ObsID 80002092004 in 2014. The detection significance is barely above 3-$\sigma$ if considered as a blind search, but the spin values and the inferred $T_0$ are very close to the values in ObsID 80002092006, which increases the reliability of this detection.
c/d show the same plot for the new detection in ObsID 90201037002, showing a detection significance well above 5-$\sigma$.
}
\label{fig:detlev}
\end{figure*}

\fref{fig:detlev} shows the statistical distribution of $Z_n^2$ values during a search for pulsations with the methods described in \sref{sec:finaldeep}. 
We executed the analysis on real data, and on (hundreds of realizations of) simulated data with the same GTIs and count rate as the original observation.
In a search containing only simulated white noise, the statistical distribution follows very closely the expected $\chi^2$ distribution \citep{buccheri_search_1983}.
This means that the folding-and-shift procedure, with a careful choice of parameters, does not alter significantly the null-hypothesis probability, that can be used to search for pulsations.

We conclude that the slight deviations from the white-noise statistical distribution found in real data are driven by the source variability and not from the analysis technique used.

To evaluate the $p$-value, in order not to increase synthetically the number of trials, we consider only the trials over $T_0$ needed to get the best detection:
if the best $T_0$ is very close to the known orbit, it means that the search over the full $T_0$ space is noise without any added value.

\subsection{Parameter files}\label{sec:parfiles}
In this Section, we write down a few useful parameter files to show the logical steps to obtain the final orbital solution.

The first step consists of fitting the two obsids with the most significant pulsations (\texttt{80002092007}, \texttt{80002092009}) and covering multiple orbits, using a spin solution with four spin derivatives, in order to constrain the orbital period and in particular the ascending node passage.
The result is the following set of timing parameters (in TEMPO2/PINT format):
\begin{verbatim}
PSR                               M82_X2_OBSID_007_009
RAJ                      9:55:42.14400000
DECJ                    69:40:26.00400000
F0                  0.7285667095459839939 1 8.741567555166416003e-08
F1              6.4250081648893346535e-11 1 4.9287115294950599775e-13
F2              2.2525682844363675388e-16 1 7.8038412567833659266e-18
F3              -5.939687798010528739e-22 1 5.0019855875541822192e-23
PEPOCH              56695.507895798600000
PLANET_SHAPIRO                          N
BINARY BT
PB                  2.5328840644740456124 1 0.00016109793561046761965
A1                     22.204213222811447 1 0.012032723626796296
ECC                                   0.0
T0                  56682.068219262420266 1 0.001018781929672351325
OM                                    0.0
\end{verbatim}

Once we do that, we can use ObsID \texttt{80002092011}, fixing $T_0$ and using the longer lever arm to better estimate the orbital period \Porb:
\begin{verbatim}
PSR                               M82_X2_OBSID_011
RAJ                      9:55:42.14400000
DECJ                    69:40:26.00400000
F0                 0.72875923701101958546 1 3.364666714127084281e-07
F1               1.189117392632877145e-10 1 6.9989856290673740436e-12
PEPOCH              56720.878600229010000
PLANET_SHAPIRO                          N
BINARY BT
PB                  2.5328387434901325451 1 4.4959856101332736067e-05
A1                     22.204213222811447 0 0.012032723626796296
ECC                                   0.0
T0                  56682.068219262420266 0 0.001018781929672351325
OM                                    0.0

\end{verbatim}
This is the solution reported as ``2014 only'' in \tref{tab:orbit}.

Finally, we can use the 2010 ObsID \texttt{90201037002} to further constrain the orbital period, obtaining:
\begin{verbatim}
PSR                                M82_X2_OBSID_90201037
RAJ                      9:55:50.20800000
DECJ                    69:40:46.99200000
F0                 0.72390599836134799007 1 7.123714905580943099e-07
PEPOCH              57641.998517226645995
PLANET_SHAPIRO                          N
BINARY BT
PB                  2.5329477619141624702 1 4.0457956366063808576e-06
A1                     22.204213222811447 0 0.012032723626796296
ECC                                   0.0
T0                  56682.068219262420266 0 0.001018781929672351325
OM                                    0.0
\end{verbatim}

\subsection{Would a high magnetic field imply an electron capture origin?}

The findings of Sections~\ref{sec:spindown} and \ref{sec:glitch} can be interpreted as signatures of a very high magnetic field.
The idea of a magnetar-scale magnetic field in a binary, however, is puzzling. 
A typical supernova explosion will result in the loss of a substantial fraction of the mass from the supernova progenitor, which will usually be the heavier object in a binary.  
When more than half the total mass in the system is lost, the binary becomes entirely unbound.  
With more modest mass loss, the binary becomes eccentric and acquires a longer orbital period than it had before the merger \citep[e.g.][]{boersma_mathematical_1961}.  
In such systems, accretion can take place only for a small range of orbital phases very close to periastron, which would suppress accretion.
Given that the typical decay timescale for a magnetar's magnetic field is of order 10000 years\footnote{\citet{igoshevHowMakeMature2018} discuss timescales up to a few million years}, but that the circularization timescale for donor stars with radiative envelopes is much longer than this, the combination of a nearly circular orbit with a magnetar-like magnetic field requires some explanation.

In principle, a very finely tuned asymmetry of the supernova explosion can ``fix'' some of the problems caused by mass loss, but these systems will have long orbital periods \citep{kalogera_formation_1998}.  
The alternative solution is to require a minimal mass loss.  
This can be done by electron capture supernova processes, which which an iron core never forms, and can result either from the collapse of intermediate mass stars \citep[e.g.][]{nomoto_evolution_1987,poelarends_supernova_2008} or accretion inducted collapse of a white dwarf in a binary \citep{ergma_accretion_1993,fryer_what_1999}.  
Some observations of accreting X-ray pulsars suggest that there are two sets of these objects, with reasonable albeit not airtight arguments that some form in core collapse supernovae and others form in electron capture supernovae \citep{pfahl_new_2002,knigge_two_2011}.  
These systems are expected to have not just lower symmetric kicks due to mass loss \citep{blaauw_origin_1961}, but also lower asymmetric kicks \citep{fryer_what_1999}.  

A few other aspects of this scenario are additionally attractive.
First, evolutionary calculations show a preference for donors of about 5-10 $M_\odot$ \citep{fragos_origin_2015}. 
The masses of stars in binaries tend to show correlations, with a flat mass ratio distribution, rather than the mass ratio distribution predicted by randomly drawing from a standard initial mass function law \citep{sana_binary_2012}.  
The direct electron capture supernovae tend to occur for progenitor masses of about 8--10 $M_\odot$ \citep{nomoto_evolution_1987}, while the white dwarfs with small enough carbon abundances that they can undergo accretion induced collapses have progenitor masses just below the range required for supernovae to occur \citep{canal_non_1976}.  
Furthermore, there are widespread suggestions that accretion-induced collapse and/or merger-induced collapse events (i.e. those in which the merger of two white dwarfs drives a collapse to a neutron star) can lead to the production of magnetars \citep[e.g.][]{usov_millisecond_1992, king_type_2001}.
Additionally, the lack of a supernova remnant associated with such a young system also favors the idea that the neutron star formed in a manner with much less mass ejection than typical core collapse supernovae.  
If indeed the magnetic field is high as suggested by the spin-down and the glitch (and as previously suggested for this source in the literature, e.g. \citealt{Tsygankov+15}, \citealt{Mushtukov+15}, \citealt{Eksi+15}, \citealt{Dallosso+15}), there is, thus, an intriguing range of indirect evidence that this system has formed via accretion-induced collapse of a white dwarf, but the evidence at the present time is far from conclusive.

\end{document}